\documentclass[aps,twocolumn,showpacs,groupedaddress,nofootinbib]{revtex4-1}
\usepackage{epsfig}
\usepackage{bm}
\usepackage{amsmath}
\usepackage{amssymb}

\newcommand{\be}{\begin{equation}}
\newcommand{\ee}{\end{equation}}
\newcommand{\ba}{\begin{eqnarray}}
\newcommand{\ea}{\end{eqnarray}}

\begin{document}

\preprint{02-02}
\title{Primakoff production of $\pi^0$, $\eta$ and $\eta'$ in the Coulomb
       field of a nucleus}
\author{Murat M. Kaskulov}
\email{murat.kaskulov@theo.physik.uni-giessen.de}
\author{Ulrich Mosel}
\affiliation{Institut f\"ur Theoretische Physik, Universit\"at Giessen,
             D-35392 Giessen, Germany}
\date{\today}

\begin{abstract}
Photoproduction of neutral pseudoscalar mesons $\pi^0,\eta(547)$ and
$\eta'(958)$  in the Coulomb field  of an atomic  nucleus is studied
using a model which describes the Primakoff and nuclear parts
of the production amplitude. At high energies the nuclear background
is dominated  by the exchange of $C$-parity odd Regge trajectories.
In the coherent production the isospin filtering makes the $\omega(782)$
a dominant trajectory.
The calculations are in agreement with $\pi^0$ data from JLAB provided the photon
shadowing and final state interactions of mesons are taken into account.
The kinematic conditions which allow to study the Primakoff effect
in $\eta$ and $\eta'$ photoproduction off nuclei are further discussed. We
also give predictions for the higher energies available at the JLAB upgrade.
\end{abstract}
\maketitle

\section{Introduction}
The coherent photoproduction of neutral pions
$$
\gamma + A \to \pi^0 + A
$$
in the Coulomb field of a nucleus was first discussed by
Primakoff~\cite{H.Primakoff} as a plausible way to measure
the $\pi^0$ life time. Indeed, at high energies the Coulomb
$\gamma\gamma^*\to\pi^0$ component gets well separated from
the background induced by nuclear interactions and shows up as
a resonant peak in the angular distributions at forward angles
- the Primakoff effect. A conversion of photons into particles
in the Coulomb field is remarkably general and can dominate in
many other reactions where the nucleus stays in its ground state
after the scattering. The Primakoff effect has been observed in
the high energy reactions involving the hadronic probes where
it was used to measure the radiative decay widths of mesons and
baryons. The most important applications of the Primakoff process
are the measurements of radiative life times of vector
$\rho^{\pm}(770)$~\cite{Berg:1980cy,Jensen:1982nf,Huston:1986wi,Capraro:1987rp},
$K^{*\pm}(890)$~\cite{Bemporad:1973pa,Berg:1980cp,Chandlee:1983hf} and axial-vector
$a_1(1260)$~\cite{Zielinski:1984au}, $b_{1}(1235)$~\cite{Collick:1985yi}
mesons in the coherent high energy  $\pi(K)\to meson$ excitation
on nuclear targets. The measurement of the $\Lambda$ to $\gamma \Sigma^0$
transition is another well known example~\cite{Devlin:1986hm,Petersen:1986fi}.
Presently the Primakoff effect is an important experimental tool to
study the physics beyond the Standard Model. For instance, the
hypothetical axions being produced in the Sun via the Primakoff
process are presently searched for using the inverse effect in
the CAST experiment at CERN~\cite{Zioutas:2004hi}. The Primakoff
physics will be studied from new perspectives with an upgrade of
JLAB to 12 GeV and at the COMPASS detector at
CERN where many reaction channels involving vector, axial-vector
and hybrid mesons will be investigated in the near future.

High energy photoproduction of pseudoscalar mesons
$\pi^0(134), \eta(547)$ and $\eta'(958)$ studied in the present work
is of special interest.  The two-photon decay widths of these mesons
are determined by the axial anomaly in QCD~\cite{Bell:1969ts,Adler:1969}.
Therefore, based on the original proposal~\cite{H.Primakoff} the
production reactions off nuclei provide a tool to measure these
quantities. This has been followed
in~\cite{Kryshkin:1969wn,Browman:1974cu,Browman:1974sj,Bemporad:1967zz}
where the $\pi^0 \to \gamma\gamma$ and $\eta\to \gamma\gamma$ decay widths
were measured. Presently a new generation of experiments~\cite{PRIMEX}
allows to measure the $\pi^0\to\gamma\gamma$ decay width via the Primakoff
effect on a few percent level. Also high precision $\eta$ and $\eta'$
experiments are planned at JLAB~\cite{etaGasp}.

However, any experimental determination of the radiative decay widths
of mesons via the Primakoff effect on nuclei has to deal with the presence
of nuclear production processes. Theoretically, it is important to control the reaction mechanisms in close vicinity of the Coulomb peak. This is because the Primakoff
production is contaminated by events which come from the coherent
and incoherent conversions in the strong field of a nucleus. Moreover,
the nuclear coherent amplitudes introduce complex phases and by interference
with the Primakoff amplitude affect the $\gamma \gamma^* \to meson$ signal.
In some cases these two mechanisms can be well separated. For instance,
at very high energies the growth of the Coulomb peak must dominate over
the Regge behavior of the strong amplitude. However, both amplitudes
necessarily interfere which makes the interpretation of the results
model dependent. The strong part of the amplitude is of much shorter
range and the nuclear conversion happens predominantly deep inside the
nucleus. Therefore, the fate of particles in the initial and final state
interactions have to be taken into account. All together these require
a model which properly deals with the nuclear background, both coherent
and incoherent, and describes the data on the same level of accuracy as
required by the experiment.

There are well formulated procedures to calculate the incoherent pion
photoproduction cross sections using {\it e.g.} the transport~\cite{Kaskulov:2008ej},
cascade~\cite{Rodrigues:2010zzc} or Glauber~\cite{Cosyn:2007er,Gevorkyan:2009mh} methods.
Morpurgo~\cite{Morpurgo1964} and   F\"aldt~\cite{Faldt1972} have laid the
theoretical basis for the treatment of the Primakoff effect on nuclei.
Both had to assume simplified expressions for the elementary strong production
amplitude. Also in the more recent approach~\cite{Gevorkyan:2009ge} which
investigates final and initial state interactions, the
elementary amplitudes were parameterized without any detailed comparison
to experimental data on the nucleon. A recent experimental
analysis~\cite{PRIMEX} has shown that agreement with experiment
could only be reached if additional parameters fixing the relative strength
and phase of the Primakoff and the nuclear coherent amplitudes from \cite{Gevorkyan:2009ge} were introduced; these parameters even depended on target-mass number.


In this work we, therefore,
develop a model for the high energy coherent production of pseudoscalar mesons off nuclei.
Starting from the high energy production off nucleons, we show, that
from the photon ($\gamma$) and Regge exchange perspectives
both the nuclear Primakoff and strong amplitudes
can be well described using the same general expressions for both components
and no free parameters such as relative strengths and phases.
We further investigate the impact of
initial and final state interactions on the reaction cross section.

The use of tagged photon beams at JLAB will allow to access $\eta\to
\gamma\gamma$ and $\eta'\to \gamma\gamma$ decay widths in the
Primakoff production off nuclei~\cite{etaGasp}. We therefore consider also the coherent
photoproduction of $\eta$ and $\eta'$ and study the kinematic
conditions which allow to separate the Primakoff peak from the
nuclear background interactions.

This work is organized as follows. In Sec.~II we discuss the radiative decays
of pseudoscalar and vector mesons in the vector dominance model.
Photoproduction of $\pi^0$,
$\eta$ and $\eta'$ mesons off nucleons with transition amplitudes derived
in Sec.~II is considered in Sec.~III.
In Sec.~IV we propose a model for the high energy coherent
production of pseudoscalar mesons off nuclei. The results are presented in Sec.~V. The
summary is given in Sec.~VI.

\section{Preliminaries}
The radiative  $\gamma\gamma$ decay widths of neutral pseudoscalar mesons
$P=\pi^0,\eta$ and $\eta'$ measured in the Primakoff process off nuclei are
tied to the description of the strong nuclear background. The latter
is dominated (at high energies) by
the exchange of vector mesons $(V)$ which interact
via the anomalous
$V\gamma\to P$ transitions.
Furthermore, at high energies
the hadronic $\gamma\leftrightarrow V$ components of the
photon in $P\to\gamma\gamma$ and $P\to V V$ transitions have to be taken into
account explicitly. In the nuclear medium this
effect  is known as shadowing, {\it i.e.}  a photon
behaves in strong interactions like a hadron~\cite{Bauer:1977iq}.
In this section we
attempt to formalize these features relevant for the present work in a common framework.

\begin{figure}[t]
\begin{center}
\includegraphics[clip=true,width=1\columnwidth,angle=0.]
{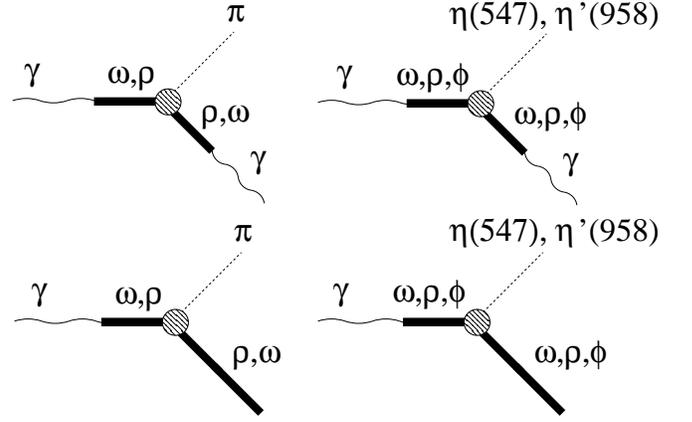}
\caption{\label{DiagramsVDM}
Diagrams describing the anomalous $\gamma + \gamma (V) \to \pi^0, \eta(547)$
and $\eta'(958)$ transition in VDM. Vector mesons $V$ include $\rho(770)$,
$\omega(782)$ and $\phi(1020)$ states.
\vspace{-0.2cm}
}
\end{center}
\end{figure}

The decays of pseudoscalar mesons $\pi^0\to\gamma\gamma$, $\eta\to\gamma\gamma$ and
$\eta'\to\gamma\gamma$
are described (at leading order) by the Wess-Zumino-Witten (WZW) term~\cite{Wess:1971yu,Witten:1983tw}, which accounts
for the anomalies in the framework of an effective theory.
The corrections to the WZW term concern
$\eta-\eta'$ mixing. In a theory which accommodates the mixing
pattern of $\eta-\eta'$ mesons  a decent
approximation to all three decays can be obtained~\cite{Leutwyler:1997yr}.
On the other hand, a common description of the anomalous $P\to \gamma\gamma$, $P\to VV$ and $P\to V\gamma$ interactions and
$\gamma\to V$ transition, where $V$ includes the $\rho^0(770)$, $\omega(782)$ and
$\phi(1020)$ vector mesons, can be realized
using the vector dominance model (VDM), see~\cite{O'Connell:1995wf} and
references therein. The corresponding VDM diagrams are shown in Fig.~\ref{DiagramsVDM}.

The anomalous $VVP$ interaction followed here reads
\be
\label{L_VVM}
\mathcal{L}_{VVP} = - \frac{N_c g^2}{32\pi^2f_{0}}
\varepsilon^{\mu\nu\alpha\beta}
\langle \partial_{\mu}V_{\nu} \partial_{\alpha}V_{\beta}P \rangle,
\ee
where
$P$ and $V_{\mu}$
describe the nonet of pseudoscalar mesons and  nonet of vector mesons,
respectively (see the Appendix).
In Eq.~(\ref{L_VVM}) $\varepsilon^{\mu\nu\alpha\beta}$ denotes the totally antisymmetric Levi-Civita tensor, $f_{0}$ is normalized to the pion decay constant
$f_{\pi}\simeq 92.4$~MeV and $N_c=3$ stands for the number of quark colors.
The coupling of the
electromagnetic field $\mathcal{A}$ to the vector mesons is given by~\cite{O'Connell:1995wf,Kawarabayashi:1966kd}
\be
\label{L_AV}
\mathcal{L}_{\gamma V} =
e \frac{m_V^2}{g}
\mathcal{A}^{\mu} \langle Q V_{\mu} \rangle,
\ee
where $Q=diag(2/3,-1/3,-1/3)$ is a quark charge matrix and $e\simeq
-\sqrt{4\pi/137}$.

Working out the traces $\langle..\rangle$ the VDM takes the form
\ba
\label{L_VVM_f}
\mathcal{L}_{VVP} = - \frac{N_c g^2}{8\pi^2f_{\pi}}
\varepsilon^{\mu\nu\alpha\beta} \Big[\partial_{\mu}\rho_{\nu}^0
  \partial_{\alpha}\omega_{\beta} \pi^0 \hspace{2cm} \nonumber \\
\left.  +
(\partial_{\mu}\rho_{\nu}^0 \partial_{\alpha}\rho_{\beta}^0 +
  \partial_{\mu}\omega_{\nu} \partial_{\alpha}\omega_{\beta})
\left[ \frac{\eta_8}{2\sqrt{3}} + \frac{\eta_0}{\sqrt{6}}  \right] \right.  \\
+ \left. \partial_{\mu}\varphi_{\nu} \partial_{\alpha}\varphi_{\beta}
\left[ \frac{-\eta_8}{\sqrt{3}} + \frac{\eta_0}{\sqrt{6}}  \right]
\right] + \cdots, \nonumber
\ea
\be
\label{L_AV_f}
\mathcal{L}_{\gamma V} =
e \frac{m_V^2}{g}
\left[\rho_{\mu}^0 + \frac{1}{3}\omega_{\mu}
-\frac{\sqrt{2}}{3} \phi_{\mu}\right] \mathcal{A}^{\mu}.
\ee
In (\ref{L_VVM_f}) only the neutral components needed in this work are shown.
As usual the observed  $\eta$ and $\eta'$ eigenstates are the
mixtures of the flavor singlet $\eta_0$ and octet $\eta_8$ states
\be
\left(
\begin{array}{c}
\eta \\
\eta'
\end{array}
\right)
=
\left(
\begin{array}{rr}
-\sin \vartheta  & \cos \vartheta \\
 \cos \vartheta  & \sin \vartheta
\end{array}
\right)
·
\left(
\begin{array}{c}
\eta_0 \\
\eta_8
\end{array}
\right),
\ee
where $\vartheta$ is the $\eta-\eta'$-mixing angle. At this point we do not
take the $\pi^0-\eta-\eta'$ isospin violating mixing into account which may become important for
a very accurate determination of the decay widths~\cite{Bernstein:2009zz}.

In the two-step $P \to VV
\to \gamma(q,\lambda)\gamma(q',\lambda')$ decay processes the transition amplitudes
are given by
\be
\label{GaGaP}
-iM_{P \to \gamma \gamma } =
 \frac{i\alpha}{\pi f_{\pi}}  c_{P}
\varepsilon^{\mu\nu\alpha\beta} q_{\mu}\epsilon_{\nu}^{\lambda\dagger}q_{\alpha}'\epsilon_{\beta}^{\lambda'\dagger},
\ee
where
$\alpha = \frac{e^2}{4\pi}$, $\epsilon^{\lambda}$ is the
polarization vector of photons ($\lambda=\pm 1$),
$c_{\pi^0}=1$, $c_{\eta}=\frac{1}{\sqrt{3}} \left(\cos\vartheta - 2\sqrt{2}
\sin\vartheta \right)$ and $c_{\eta'}=\frac{1}{\sqrt{3}} \left(\sin\vartheta + 2\sqrt{2}
\cos\vartheta \right)$.
Using (\ref{GaGaP}) the $\gamma\gamma$ decay widths of neutral pseudoscalars are given by
\be
\label{GammaPi}
\Gamma_{P \to \gamma\gamma} = \frac{\alpha^2
  }{64 \pi^3 f_{\pi}^2} m_{P}^3 c^2_{P} \, .
\ee
The measured values are well described using $\vartheta\simeq -16^{\circ}$
resulting in $\Gamma_{\pi\to \gamma\gamma} = 7.73$~eV, $\Gamma_{\eta \to
  \gamma\gamma} = 0.52$~keV and $\Gamma_{\eta' \to\gamma\gamma} = 5.48$~keV.
The masses of mesons are~\cite{PDG} $m_{\pi^0}=134.9$~MeV, $m_{\eta}=547.9$~MeV and
$m_{\eta'}=957.8$~MeV.

The experimental $\gamma\gamma$
decay width of $\pi^0$ is~\cite{PDG}
\be
\label{PIGGPDG}
\Gamma_{\pi^0 \to \gamma\gamma}^{exp} = (7.74\pm 0.55)~\mbox{eV},
\ee
as required by the anomaly.
Note that the corresponding decay width measured via the Primakoff effect
$\Gamma_{\pi^0 \to \gamma\gamma} = (7.82\pm 0.14)~\mbox{eV}$~\cite{PRIMEX}
agrees with~(\ref{PIGGPDG}). However, in the $\eta$ chanel the $\gamma\gamma$ decay
width $\Gamma_{\eta \to \gamma\gamma} =  (0.32 \pm  0.046 )~\mbox{keV}$ measured at Cornell~\cite{Browman:1974sj} when using the Primakoff effect\footnote{In
    Ref.~\cite{Sibirtsev:2010cj} $\Gamma_{\eta \to \gamma\gamma}$ has been
    fitted to the Primakoff effect off protons. The value of  $\Gamma_{\eta \to
     \gamma\gamma}$ from this fit is $\sim 50$\% bigger than from $e^+e^-$
    experiments, Eq.~(\ref{GammaEta}). However, the authors
    of~\cite{Sibirtsev:2010cj} miss the factor $e^2$ in their expressions for the Primakoff cross section
    which makes the Coulomb components of~\cite{Sibirtsev:2010cj}
    and~\cite{Sibirtsev:2009kw} too large by one order of magnitude.}
 is at variance with the measurements via the QED production in
$e^+e^-$ collisions~\cite{PDG} where
\be
\label{GammaEta}
\Gamma_{\eta \to \gamma\gamma}^{exp} =  (0.51 \pm  0.026)~\mbox{keV}.
\ee
As for $\eta'\to\gamma\gamma$ decay the width was  measured in the $e^+e^-$
collisions with the value of~\cite{PDG}
\be
\label{GammaEtaP}
\Gamma_{\eta' \to \gamma\gamma}^{exp} = (4.28 \pm 0.19)~\mbox{keV}.
\ee

In the Primakoff production one of two photons is off-mass-shell. In VDM the
half off-shell $\gamma(q)\gamma^*(l)\to P(k)$ vertices to be used in the following read
\ba
\label{MgP}
-iM_{\gamma\gamma^*\to P}^{\beta} &=&
 \frac{i\alpha}{\pi f_{\pi}}
\varepsilon^{\mu\nu\alpha\beta} q_{\mu}
\epsilon_{\nu}^{\lambda}l_{\alpha} \,
  F_{\gamma\gamma^*}^{P}(l^2),
\ea
where
\ba
\label{MgPF1}
F_{\gamma\gamma^*}^{\pi}(l^2) = \frac{1}{2} \left[F_{\rho}(l^2)+F_{\omega}(l^2)\right],
\ea
\ba
\label{MgPF2}
F_{\gamma\gamma^*}^{\eta}(l^2) &=& \sqrt{3} \left[ \left[
  F_{\rho}(l^2) + \frac{1}{9}F_{\omega}(l^2) \right] \right.
 \left. \left(\frac{\cos\vartheta}{2}-\frac{\sin\vartheta}{\sqrt{2}}\right) \right. \nonumber\\
&& \left.
- \frac{2}{9} F_{\phi}(l^2)
\left(\cos\vartheta+\frac{\sin\vartheta}{\sqrt{2}}\right)\right],
\ea
\ba
\label{MgPF3}
F_{\gamma\gamma^*}^{\eta'}(l^2) &=& \sqrt{3}
\left[ \left[
  F_{\rho}(l^2) + \frac{1}{9}F_{\omega}(l^2) \right]
  \left(\frac{\cos\vartheta}{\sqrt{2}}+\frac{\sin\vartheta}{2}\right)
\right. \nonumber\\
&& \left.
+ \frac{2}{9} F_{\phi}(l^2)
\left(\frac{\cos\vartheta}{\sqrt{2}}-\sin\vartheta\right)\right].
\ea
with VDM (form) factors $F_V(l^2) = \frac{m_V^2}{m_V^2-l^2-i0^+}$. For the
calculations of form factors in an improved VDM framework see \cite{Terschluesen:2010ik}.

The transition amplitude $\gamma(q) V(l) \to P(k)$ relevant for our
purpose can be derived from the Lagrangian Eq.~(\ref{L_VVM_f}) and from the
$\gamma V$ transitions of the non-anomalous Lagrangian Eq.~(\ref{L_AV_f}).
It takes the form
\be
\label{MGVP}
-i M_{\gamma V \to P} = - i e G_{\gamma VP}
\varepsilon^{\mu\nu\alpha\beta} q_{\mu} \epsilon_{\nu}^{\lambda}l_{\alpha}\epsilon_{\beta}^{\lambda'(V)},
\ee
where $G_{\gamma VP} = \frac{g }{8\pi^2f_{\pi}}b_{(P)}^{\gamma(V)}$,
$\epsilon^{\lambda'(V)}$ is the polarization vector of $V$
with ${\lambda'=0,\pm 1}$ and the coefficients $b_{(P)}^{\gamma (V)}=b_{(\pi,\eta,\eta')}^{\gamma (\rho,\omega,\phi)}$ are given by
\be
\label{bcoeff}
\begin{array}{ccc|c}
\gamma\rho  & \gamma\omega & \gamma\phi & \\
\hline
1 & 3 & 0 & \pi \\
{\sqrt{3}}{
\left(\cos\vartheta-\sqrt{2}\sin\vartheta \right)} &
\frac{\left(\cos\vartheta-\sqrt{2}\sin\vartheta\right)
}{\sqrt{3}} &
\frac{\left(2 \sqrt{2}\cos\vartheta+2\sin\vartheta\right)}{\sqrt{3}} &
\eta
\\
{\sqrt{3}}{
\left(\sin\vartheta+\sqrt{2}\cos\vartheta\right)} &
\frac{\left(\sin\vartheta+\sqrt{2}\cos\vartheta\right)
}{\sqrt{3}} &
\frac{\left(2\sqrt{2}\sin\vartheta-2\cos\vartheta\right)}{\sqrt{3}}
& \eta'
\end{array} \nonumber
\ee

The VDM coupling $g$ is supposed
to be universal.
The value of
$g\simeq 6$ is determined from the $\rho^0\to\pi^+\pi^-$ decay width
$\Gamma_{\rho^0\to \pi^+\pi^-} = \frac{g^2}{48\pi} m_{\rho}
\left(1-{4m_{\pi}^2}/{m_{\rho}^2} \right)^{3/2}$ and approximates well
the transition rates in~(\ref{MGVP}).
For instance, in the reaction channels $\gamma \rho(\omega) \to \pi^0,\eta,\eta'$ needed in the following
the corresponding radiative decay widths
$
\Gamma_{V\to\gamma P} = \frac{\alpha}{24}
\frac{G_{\gamma V P}^2}{m_{V}^3}
\left({m_{V}^2} - {m_{P}^2}\right)^3$ and/or  $\Gamma_{P\to\gamma V} = \frac{\alpha}{8}
\frac{G_{\gamma V P}^2}{m_{P}^3}
\left({m_{P}^2} - {m_{V}^2}\right)^3$ calculated using the universality of
$g$ and~(\ref{MGVP}) agree well with~\cite{PDG}.
The coupling $g$  will also show up in the interaction of
$\rho$-meson with nucleons.
In this form the anomalous $\gamma V \to P$ and strong
$\rho NN$ interactions get correlated
reducing largely the number of free parameters.

In the photoproduction of $P$ the vector mesons are exchanged in the
$t$-channel.
The corresponding transition vertices which will enter the production
amplitude off nucleons and nuclei read
\be
\label{MVPG}
-i M_{\gamma V^* \to P}^{\beta} = - i e G_{\gamma VP}
\varepsilon^{\mu\nu\alpha\beta} q_{\mu} \epsilon_{\nu}^{\lambda}l_{\alpha}.
\ee

\section{Photoproduction off nucleons}
The production of charged pions off nucleons with high energy real and
virtual photons has been studied in~\cite{Kaskulov:2010kf}. In the
neutral $\pi^0$ channel the Primakoff effect together with exchanges of $C$-parity
odd vector $\rho(770),\omega(782)$ and axial-vector $b_1(1235)$ and $h_1(1170)$
Regge trajectories
has been calculated~\cite{KM2}. As a novel feature the contributions of $s$- and
$u$-channel nucleon resonances were investigated. The latter were described using
the Bloom-Gilman connection between the exclusive and (deep inelastic) inclusive reactions.
The resonances are effective in the
off-forward region around the diffractive dip and in the deeply virtual (high
$Q^2$) electroproduction where the quarks are the relevant degrees of freedom~\cite{Kaskulov:2008xc}.

At the real photon point the high energy forward production of $\pi^0$ is
dominated by the $t$-channel exchange of mesons, with
$\omega$ and $\rho$ being the dominant Regge trajectories. For the $\eta$ and $\eta'$ channels
we assume the dominance of these exchange contributions
as well. In the model of~\cite{Kaskulov:2010kf,KM2} 
the sum of $s$- and $u$-channel contributions
cancel at the forward angles. 
Also the contributions of axial-vector mesons are
small and can be readily neglected. A detailed comparison with data on the nucleon has
been performed in \cite{Kaskulov:2010kf,KM2}.
The dominance of $\omega$- and $\rho$-exchanges  at
forward angles makes the model similar (up to the
Regge phase of the $\rho$-exchange) to that proposed by Laget in~\cite{Laget:2005be}.
However, contrary to~\cite{Laget:2005be} we rely on a VDM
framework which resolves the hadronic structure of the anomalous vertex needed
in a consistent treatment of shadowing and allows to
fix the relative sign between the Primakoff and strong amplitudes. The VDM
form factors further introduce  an additional $t$ dependence;
they become important in the production of heavy $\eta$ and $\eta'$
mesons which require sizable momentum transfer to the target already at
forward angles.

In the Primakoff production of $\pi^0,\eta$ and $\eta'$ off protons the
amplitude takes the form
\ba
\label{T_Pr}
-i{T}^{\gamma}_{\gamma p\to Pp} &=&   - i |e|
\, M_{\gamma\gamma^*\to P}^{\mu} \, D_{\mu\nu}^{\gamma}(l)
\\ &&\times \bar{u}_{s'}(p')
\left[F_{1}^{p} \gamma^{\nu} -
  F_{2}^{p} \frac{i\sigma^{\nu\sigma}l_{\sigma}}{2m_N} \right] u_s(p), \nonumber
\ea
where $l=p-p'$, $D_{\mu\nu}^{\gamma}$ is the photon propagator
\be
\label{G_FP}
iD_{\mu\nu}^{\gamma}(l) = \left[-g_{\mu\nu} +\frac{l_{\mu}l_{\nu}}{l^2}\right] \frac{i}{l^2+i0^+},
\ee
and $F_{1}^p(F_{2}^p)$ stands for the Dirac (Pauli) form factor. In
Eq.~(\ref{T_Pr}) the anomalous VDM
transition vertex $M_{\gamma\gamma^*\to P}^{\mu}$ is given by (\ref{MgP}). At forward
angles the
contribution of the tensor term $\propto F_{2}^p$ is marginal  and can be readily neglected.
The parametrization of $F_{1}^p$ used here is
given in~\cite{Kaskulov:2010kf} and follows the results
of~\cite{Kaskulov:2003wh}. However, one could safely take $F_{1}^p=1$ without
much impact on the results.

\begin{figure}[t]
\begin{center}
\includegraphics[clip=true,width=1\columnwidth,angle=0.]
{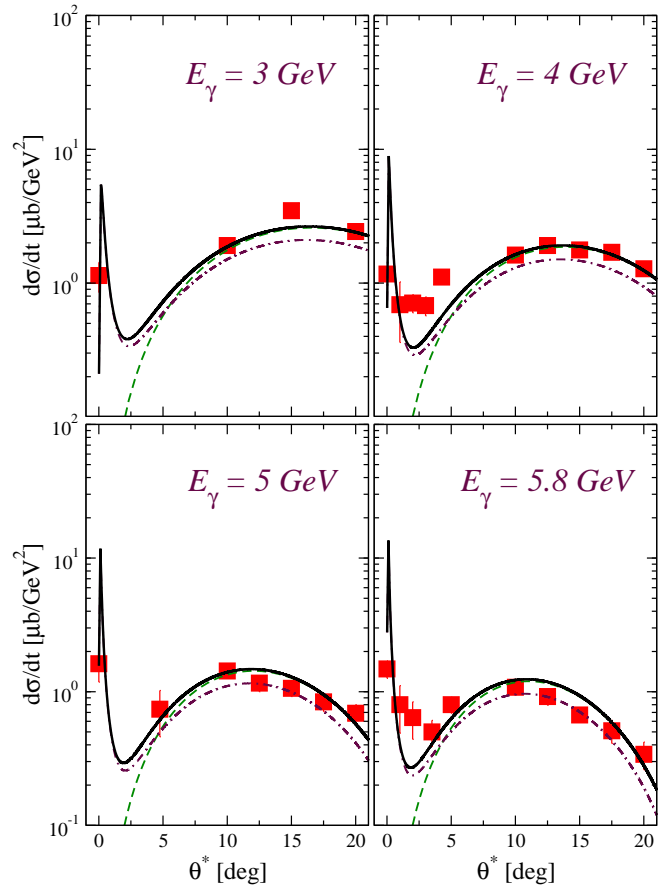}
\caption{\label{PrimPiPrN}
Differential cross section $d\sigma/dt$ as a function of the center of mass
scattering angle $\theta^*$ in the reaction $p(\gamma,\pi^0)p$. The solid
curves describe the model results with Primakoff and Regge exchange
contributions. The dashed curves are the Regge exchange contributions.
The dash-dotted curves describe the cross section without the
$\rho$-exchange. The experimental data are from Ref.~\cite{Braunschweig:1970dp}.
\vspace{-0.7cm}
}
\end{center}
\end{figure}

The strong amplitude which describes the photoproduction
of pseudoscalar mesons off nucleons by exchange of
vector mesons is given by
\ba
-i{T}^{V}_{\gamma N\to PN} &=&   - i
G_{VNN} \, M_{\gamma V^{*}\to P}^{\mu} \, D_{\mu\nu}^{V}(l)
\\ &&\times \bar{u}_{s'}(p')
\left[
\gamma^{\nu} -
\kappa_V \frac{i\sigma^{\nu\sigma}l_{\sigma}}{2m_N} \right] u_s(p). \nonumber
\ea
Here $M_{\gamma V^{*}\to P}^{\mu}$ is given by (\ref{MVPG}) and $D_{\mu\nu}^{V}$ denotes the Feynman propagator of a $V$-meson
\be
\label{V_FP}
iD_{\mu\nu}^{V}(l) = \left[-g_{\mu\nu} +\frac{l_{\mu}l_{\nu}}{m_V^2}\right] \frac{i}{l^2-m_V^2+i0^+}.
\ee
The interactions of the singlet $V_0^{\mu}$ and of the octet $V_8^{\mu}$ of vector mesons
with baryons (octet) $B$ are of the form
\be
\label{VBBvector}
v_F \langle \bar{B} \gamma_{\mu} [V^{\mu}_8,B] \rangle + v_D
\langle \bar{B} \gamma_{\mu} \{V^{\mu}_8,B\}\rangle  + v_S \langle \bar{B} \gamma_{\mu}B
\rangle
V_0^{\mu}
\ee
where $[..](\{..\})$  commutes (anti-commutes) SU(3) flavor states. Imposing the OZI rule and an ideal mixture between the
singlet and the $I=0$ of the octet one gets
$
v_S = \sqrt{\frac{2}{3}}(3v_F -v_D).
$
Then all the other coupling constants are expressed in terms of $F$ and $D$
coupling constants. The couplings of the $\rho^0$ and $\omega$ to nucleons are
given by $G_{\rho}=G_{\rho pp}=v_F+v_D$ and $G_{\omega}=G_{\omega
  pp}=G_{\omega nn}=3v_F-v_D$, respectively.  For the tensor $VNN$ coupling
one uses a similar flavor structure with similar $F/D$ relations between
constants. The $\rho NN$ coupling is isovector, $G_{\rho nn} =
-G_{\rho pp}$, and is approximately
universal~\cite{Kaskulov:2010kf,Laget:2005be}, that is  $G_{\rho}\simeq g/2$, where $g$ enters the VDM equations. Therefore, we
assume the same value of $g$ in the $\rho NN$ and in the $\gamma\rho(\omega)
P$ vertices. The flavor symmetry does not correlate the $\omega
NN$ and $\rho NN$ couplings and we use $G_{\omega}$ as a fit parameter.
The anomalous $\rho NN$ coupling is $\kappa_{\rho}=6.1$ as in the
charged pion photoproduction~\cite{Kaskulov:2010kf,Laget:2005be}. The $\omega NN$
tensor coupling is known to be very small $\kappa_{\omega}\simeq 0$.

In the photoproduction amplitude we have the contributions from the $\gamma$
and from the $\rho$ and $\omega$ exchanges
\be
{T}_{\gamma N\to PN} = \sum\limits_{m=\gamma,\omega,\rho}{T}^{m}_{\gamma N\to PN}.
\ee
However, at high energies a single pole approximation to the meson-exchange
currents is not sufficient; the corresponding Feynman diagrams diverge.
To regulate the high energy behavior we account for the higher mass
(spin) excitations of exchanged vector mesons using the Regge propagator
($D^V_{\mu\nu}\to R^V_{\mu\nu}$)
\ba
\label{ReggeProp}
iR^V_{\mu\nu}(l) &=&
i\left(-g_{\mu\nu}+\frac{l_{\mu}l_{\nu}}{m_{V}^2} \right)
 \left[\frac{1 - e^{-i\pi\alpha_{V}(l^2)}}{2}\right]
\nonumber \\
&\times&
\left(- \alpha'_{V} \right) \Gamma[1-\alpha_{V}(l^2)]
e^{\ln(\alpha'_{V}s)(\alpha_{V}(l^2)-1)}, ~~~
\ea
where $s=(p+q)^2$ and $\alpha_{V}$ is the Regge trajectory of vector mesons
\be
\alpha_{V}(l^2) = \alpha^0_V + \alpha_V' l^2.
\ee
The $\Gamma$-function in Eq.~(\ref{ReggeProp}) contains the pole propagator
$\sim 1/\sin(\pi\alpha_{V}(l^2))$ but no zeros and the amplitude's zeros
only occur through the factor $1 - e^{-i\pi\alpha_{V}(l^2)}$. In the limit
$l^2\to m_V^2$ Eq.~(\ref{ReggeProp}) is reduced to the standard Feynman
propagator, Eq.~(\ref{V_FP}). For the $\omega$-trajectory we use
$\alpha_{\omega}^0= 0.4$ and $\alpha_{\omega}'=0.85$~GeV$^{-2}$ from the
fit to forward and off-forward $\pi^0$ data~\cite{KM2}.
The Regge trajectory of $\rho$ is described by $\alpha_{\rho}^0 = 0.53$
and $\alpha_{\rho}'=0.85$~GeV$^{-2}$ as in the $\pi^{\pm}$ electroproduction~\cite{Kaskulov:2010kf}.

\begin{figure}[t]
\begin{center}
\includegraphics[clip=true,width=1\columnwidth,angle=0.]
{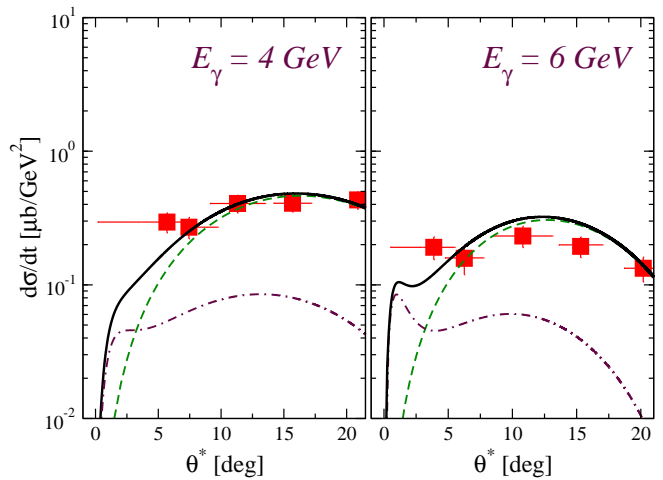}
\caption{\label{PrimEtaN}
Differential cross section $d\sigma/dt$ as a function of
the center of mass scattering angle $\theta^*$ in the reaction $p(\gamma,\eta)p$ at forward angles.
The notations for the curves are the same as in Fig.~\ref{PrimPiPrN}. The
experimental data are from Ref.~\cite{Braunschweig:1970jb}.
\vspace{-0.2cm}
}
\end{center}
\end{figure}

The results for $d\sigma/dt$ in the reaction $p(\gamma,\pi^0)p$ at forward center of
mass scattering angles $\theta^*$ are shown in Fig.~\ref{PrimPiPrN}.
The experimental data are from Ref.~\cite{Braunschweig:1970dp} and correspond
to the energies of incoming photons $\nu=3, 4, 5$ and $5.8$~GeV in the laboratory.
In these calculations we used the value of $G_{\omega}=19$. Note that in the
fit of low energy $NN$ models the coupling $G_{\omega}$ is smaller
$10 < G_{\omega}^2/4\pi < 20$~\cite{Machleidt:1987hj}. However, the latter
values are always tied to the additional $\omega$-nucleon form factors which are
replaced here by the Reggeized propagators. Our choice of $G_{\omega}$ is
necessarily not the optimal one since it also depends on the parameters of the
$\omega$-Regge trajectory used here.
The solid curves in Fig.~\ref{PrimPiPrN} describe the model results with the Primakoff ($\gamma$) and
Regge ($\omega,\rho$) exchange
contributions. The strong rise of the cross section at extreme forward angles is due to the Coulomb
component (Primakoff effect).
In the Regge-exchange part (dashed curves) the dominant contribution comes from the
$\omega$-exchange. The dash-dotted curves describe the results without
the $\rho$-Regge exchange.

\begin{figure}[t]
\begin{center}
\includegraphics[clip=true,width=1\columnwidth,angle=0.]
{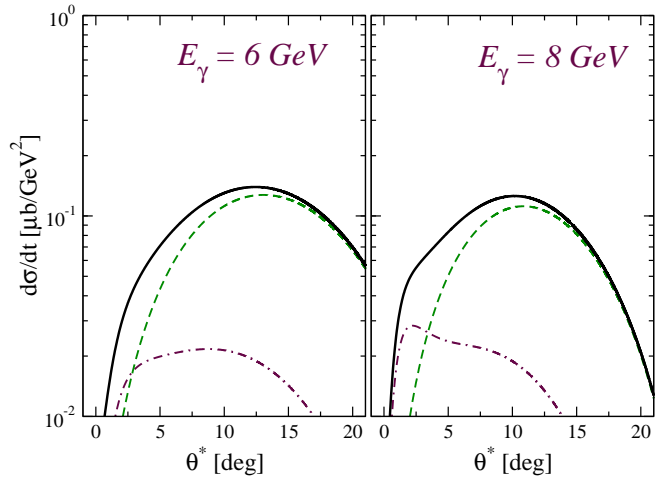}
\caption{\label{PrimEtaPrN}
Differential cross section $d\sigma/dt$ as a function of the center of
mass scattering angle $\theta^*$
in the reaction $p(\gamma,\eta')p$ at forward angles.
The notations for the curves are the same as in Fig.~\ref{PrimPiPrN}.
\vspace{-0.2cm}
}
\end{center}
\end{figure}

\begin{figure*}[t]
\begin{center}
\includegraphics[clip=true,width=2\columnwidth,angle=0.]
{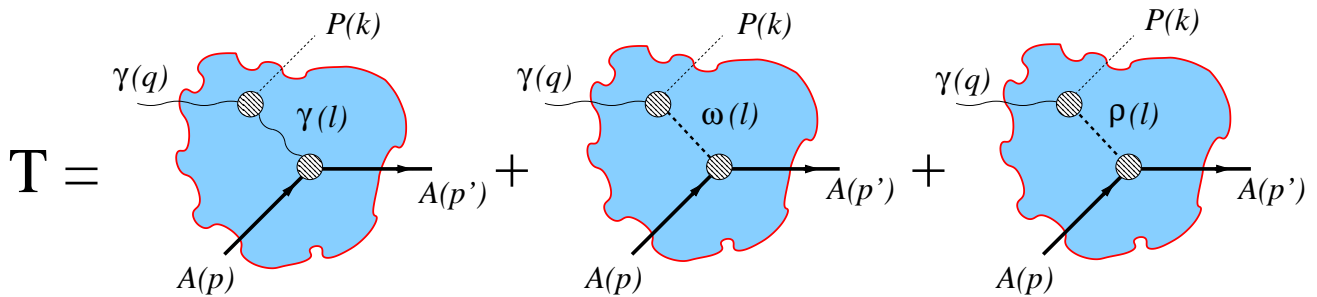}
\caption{\label{DiagramPrimakoff}
The diagrams describing the Primakoff effect (left diagram) and nuclear background
$\omega$- and $\rho$-Regge exchange amplitudes in high energy coherent $P=\pi^0$, $\eta$
and $\eta'$
photoproduction off nuclei.
 The shaded regions describe a possible conversion in the nuclear medium.
In this case the outgoing mesons experience FSI. The incoming photons
get shadowed in ISI.
\vspace{-0.7cm}
}
\end{center}
\end{figure*}

The results for $d\sigma/dt$ in the $\eta$ photoproduction off protons are shown in Fig.~\ref{PrimEtaN}.
Here the model calculations (solid curves) at $\nu=4$ and $6$~GeV are compared with
forward data from Ref.~\cite{Braunschweig:1970jb}.
The dashed curves describe the Regge-exchange contributions.
In the reaction $(\gamma,\eta)$ off nucleons the exchange of
the $\rho$-Regge trajectory dominates the reaction mechanism.
Furthermore, the VDM coupling $g$ and $\eta-\eta'$ mixing angle
$\vartheta$ essentially determine the magnitude
of the cross section. The same mixing angle enters the Coulomb part of the
amplitude and therefore determines the $\Gamma_{\eta\to\gamma\gamma}$ decay
width. As one can see, at these energies the Primakoff effect in
$(\gamma,\eta)$ is barely visible on top of
the large hadronic background. The available high energy $(\gamma,\eta)$ data off protons
do not allow any reliable extraction of $\Gamma_{\eta \to \gamma\gamma}$ at forward angles.

However, in the coherent production off nuclei the isovector $\rho$-exchange
is filtered out, {\it i.e.} in  nuclei
the $\rho$-induced $\gamma\to\eta$ conversion off neutrons adds up
destructively to the production off protons. Therefore, being effective
in the elementary production off nucleons the $\rho$-exchange
gets largely reduced in coherent reactions off nuclei due to isospin filtering.
The dash-dotted curves in Fig.~\ref{PrimEtaN} describe the model calculations with $\gamma$
and $\omega$ exchange contributions only. In this case the Primakoff peak gets
much more pronounced on top of the $\omega$-exchange background. In fact, this
is a situation which is realized in the nuclear coherent production of  $\eta$
mesons.

The Primakoff effect in the photoproduction of $\eta'$ off protons at $\nu=6$
and $8$~GeV in the laboratory is shown in
Fig.~\ref{PrimEtaPrN}.  Here the notations for the curves are the same as in
Fig.~\ref{PrimEtaN}. As one can see, the results are qualitatively similar to the $\eta$
case. The differential  cross section beyond the
Coulomb region is again dominated by the exchange of the $\rho$-Regge
trajectory. Because of the higher $\eta'$ mass  the energy of the photons
should also be
much higher to separate the Primakoff component from the contribution of meson-exchange
currents.

\section{Coherent production off nuclei}
The diagrams in
Fig.~\ref{DiagramPrimakoff} describe the
Primakoff effect and the $\omega$ and $\rho$ exchange contributions
in the coherent $\pi^0$, $\eta$
and $\eta'$
photoproduction off nuclei.
In the coherent reactions the residual nucleus remains in the
ground state. The shaded regions describe the
$\gamma \to meson$ conversion in the
nuclear medium. In this case the outgoing mesons experience final state
interactions (FSI). The incoming photons get shadowed - initial state
interactions (ISI).

The differential cross section in the reaction $\gamma(q) +
A(p) \to \pi^0(k) + A(p')$ describing the bare Primakoff effect (without
FSI and ISI)  is given by~\cite{Morpurgo1964,Hadjimichael:1989ks}
\ba
\label{PrimStandF}
\frac{d\sigma}{d\Omega} &=& \Gamma_{\pi^0\to \gamma\gamma} \frac{8\alpha
  Z^2}{m_{\pi}^3}
 \frac{|\vec{k}|^3 \nu}{t^2} \, \mathcal{F}_A^2(t) \, \sin^{2}\theta
\ea
where $t=(k-q)^2=(p-p')^2$. This formula is based on the assumption
that the $\gamma\to\pi^0$ conversion
$\sim ({\bf
  E} \cdot {\bf
  B})$ takes place in the electric field ${\bf
  E}$ of infinitely heavy spinless ($J=0$) nucleus. The form
factor $\mathcal{F}_A$ is related to the classical field ${\bf
  E}$ by the Fourier transform of the nuclear charge ($Z$) density
$\rho_Z(\vec{x})$~\cite{Hadjimichael:1989ks}.

Alternatively, one can derive Eq.~(\ref{PrimStandF}) using a field theoretical description.
Neglect again the spin of a nucleus. Then the  nuclear matrix element of the
nucleon vector current is
\ba
\label{PrimAvertex}
-i \langle A(p') |  |e| \overline{\psi} \gamma^{\mu} \psi | A(p) \rangle \hspace{3cm}
 \\ =
-i\frac{|e| Z e^{-i(p-p')x}}{\sqrt{2 E}\sqrt{2 E'}}  (p+p')^{\mu}
\mathcal{F}_{A}(t). \nonumber
\ea
This is just a conserved current of a spinless state where
the internal substructure is realized in the momentum space
form factor $\mathcal{F}_{A}$. Indeed, using Eq.~(\ref{PrimAvertex}) as the
$\gamma AA$ vertex and using the anomalous $\gamma\gamma\to \pi^0$ transition amplitude, Eq.~(\ref{GaGaP}),
the $\gamma$-exchange Feynman diagram in Fig.~\ref{DiagramPrimakoff}  results (for $m_A\to\infty$) in Eq.~(\ref{PrimStandF}).

The $\omega AA$ and $\rho AA$ interactions take essentially the same
form as the $\gamma AA$ interaction. For instance, the $\omega AA$
interaction is obtained replacing the electromagnetic coupling
$G_{\gamma A}= |e|Z$ by the strong coupling $G_{\omega A} = A
G_{\omega}$. Note that, by this we assume that the hadronic (matter) and
electromagnetic form factor distributions are essentially the same.
The $\omega$ couples to the isoscalar-vector  currents inside the
nucleus and all the $A$ nucleons contribute constructively to the
$\omega A$ interaction.

The $\rho$-meson couples to the
isovector currents: its contribution is proportional to the difference between
the number of protons and neutrons $G_{\rho A} =(2Z-A)G_{\rho}$ and therefore
is largely suppressed. An exact $\rho$  filtering is realized when $Z=N$.
At the forward angles studied here the $\sim \kappa_{\rho} \sigma^{\mu\nu} $
term can be merely neglected. On the other hand, the same tensor coupling
is additionally suppressed in ground state nuclei. The corresponding nuclear
tensor (helicity flip) currents $-i \langle A(p') |  |e| \overline{\psi} \sigma^{\mu\nu} \psi |
A(p) \rangle $ are marginal. This is known from the elastic electron
scattering off nuclei and also from existing microscopic models for coherent
$\pi^0$ and $\eta$ photoproduction off
nuclei~\cite{Piekarewicz:1997vj,Peters:1998hm,Peters:1998mb,Kaskulov:2001hk}.

\subsection{General production amplitude with final state interactions}

In this section  we account for the fate of particles in
the final state interactions. It follows 
that the interactions of $m=\gamma,\omega$ and $\rho$
with a nucleus can be described by the Lagrangian
\ba
\label{LeffPrim}
\mathcal{L}_{m A}(z) = -i G_{m A} (C^{\dagger}(z) \partial_{\mu} C(z) -
(\partial_{\mu} C^{\dagger}(z))  C(z)) \nonumber \\
\times \int d^4 y  F_A (y-z) \mathcal{V}^{\mu}_m(y), \hspace{0.5cm}
\ea
where the profile function $F_A$ of a nucleus
makes the interactions of the fields $\mathcal{V}^{\mu}_{m}$ ($\mathcal{V}^{\mu}_{m}=\mathcal{A}^{\mu},
\omega^{\mu}$ or $\rho^{0\mu}$) non-local. $F_A$ is normalized as follows $\int d^4 y
F_A(y) = 1$, where $F_A=\delta^4(y-z)$ corresponds to a point like
nucleus. The Fourier transform of $F_A$ is a form factor introduced in
Eq.~(\ref{PrimAvertex}). In (\ref{LeffPrim}) $C^{\dagger}(C)$ creates (annihilates) a nucleus with
$C^{\dagger}(z)= \int \frac{d^3 \vec{k}}{\sqrt{V}}\frac{a^{\dagger}_{k}}{\sqrt{2 E_{k}}}
e^{ikz}$ with {\it e.g.} $V=(2\pi)^3$.

Using the VDM transition vertices, see Eqs.~(\ref{MgPF1}) and (\ref{MVPG}),
and using Eq.~(\ref{LeffPrim})
the amplitude describing the reaction
$\gamma(q,\lambda) + A(p) \to P(k) + A(p')$ is given by
\ba
\label{TGA}
-iT^{\lambda} &=& -i
\int d^4 x \int d^4 y \int d^4 z \int \frac{d^4 l}{(2\pi)^4} F_A(y-z)
\nonumber \\ && \times \left[ \sum\limits_{m=\gamma,\omega,\rho}
   d^{(m)}_P(l) \right] \varepsilon^{\mu\nu\alpha\beta}q_{\mu}\epsilon_{\nu}^{\lambda}
  l_{\alpha}(p+p')_{\beta}  \\
&& \times
 e^{-i(p-p')z} e^{-il(x-y)} e^{-i(q-k)x}
\chi^{(-)*}_{P}(\vec{x}-\vec{z}\,). \nonumber
\ea
Here $d^{(m)}_P(l)$ absorbs the propagator functions and coupling
constants defined as
\be
\label{d_gamma}
d^{(\gamma)}_{P}(l) = \frac{|e| }{\pi f_{\pi}} \frac{\alpha Z}{l^2+i0^+}  F_{\gamma\gamma^*}^{P}(l^2),
\ee
\ba
\label{d_omega}
d^{(\omega)}_{P}(l) &=& |e| A G_{\omega} G_{\gamma\omega P}\left[\frac{1 -
    e^{-i\pi\alpha_{\omega}(l^2)}}{2}\right]  \\
&& \times \left(- \alpha'_{\omega} \right) \Gamma[1-\alpha_{\omega}(l^2)]
e^{\ln(\alpha'_{\omega}\bar{s})(\alpha_{\omega}(l^2)-1)}, \nonumber
\ea
\ba
\label{d_rho}
d^{(\rho)}_{P}(l) &=& |e| (2Z-A) G_{\rho} G_{\gamma\rho P}\left[\frac{1 -
    e^{-i\pi\alpha_{\rho}(l^2)}}{2}\right]  \\
&& \times \left(- \alpha'_{\rho} \right) \Gamma[1-\alpha_{\rho}(l^2)]
e^{\ln(\alpha'_{\rho}\bar{s})(\alpha_{\rho}(l^2)-1)}, \nonumber
\ea
for $\gamma$, $\omega$ and $\rho$ exchange currents, respectively. The VDM form factors $F_{\gamma\gamma^*}^{P}$
in Eq.~(\ref{d_gamma}) are defined in~(\ref{MgPF1}-\ref{MgPF3}). Since $\nu$
is much bigger than the average Fermi momentum of nucleons
one can take $\bar{s}=m_N^2+2\nu m_N$ in (\ref{d_omega}) and (\ref{d_rho}).

Suppose, the $\gamma \gamma\to P$ or $\gamma V \to P$ conversions occur inside the
nucleus. Then the outgoing meson wave acquires
an additional eikonal phase $\chi^{(-)*}_{P}$, see Eq.~(\ref{DFmeson}) bellow,
due to FSI.
Making use of the distorted wave
approximation the meson wave function in (\ref{TGA}) reads
\ba
\varphi^{(-)*}_{P} &=& e^{ikx} \chi^{(-)*}_P.
\ea
In a plane wave approximation $\chi^{(-)*}_P=1$.

At first, in Eq.~(\ref{TGA}) one has to integrate out the variable $z$  which describes
the center of mass motion of a nucleus. Introduce new variables
$ \xi =  y-z~~\mbox{and}~~ \zeta = x-z$
with
$d^4xd^4yd^4z=  d^4\xi d^4\zeta d^4z.$
Then one gets
\ba
\label{TcmRemove}
-iT^{\lambda} &=& -i  \int d^4z
e^{-i(p-p'+q-k)z} \int  \frac{d^4 l}{(2\pi)^4}
\left[ \sum\limits_{m=\gamma,\omega,\rho}
   d^{(m)}_P(l) \right] \nonumber \\
&\times&   \varepsilon^{\mu\nu\alpha\beta}q_{\mu}\epsilon_{\nu}^{\lambda}
l_{\alpha}(p+p')_{\beta} \int d^4 \xi  e^{il\xi} F_A(\xi)
\\
&\times&
\int d^4 \zeta e^{-i(l-k+q)\zeta} \chi^{(-)*}_{P}(\vec{\zeta}\,), \nonumber
\ea
where $\int d^4z
e^{-i(p-p'+q-k)z}=(2\pi)^4\delta^4(p-p'+q-k)$ describes the
four-momentum conservation in the reaction.
The Fourier integral
$
\int d^4 \xi  e^{il\xi} F_A(\xi)  = \mathcal{F}_A(l)
$
describes the nuclear form factor in the momentum space, see Eq.~(\ref{PrimAvertex}).
By this Eq.~(\ref{TcmRemove}) takes the form
\ba
\label{Amplitude5}
-iT^{\lambda} &=& -i  (2\pi)^4\delta^4(p-p'+q-k)
\nonumber \\
 &\times&   \int \frac{d^4{l}}{(2\pi)^4} \left[ \sum\limits_{m=\gamma,\omega,\rho}
   d^{(m)}_P(l) \right] \mathcal{F}_A({l})
 \\
&\times&  \varepsilon^{\mu\nu\alpha\beta}q_{\mu}\epsilon_{\nu}^{\lambda}
  l_{\alpha}(p+p')_{\beta}
\int d^4{\zeta}  e^{-i(l-k+q)\zeta} \chi^{(-)*}_{P}(\vec{\zeta}\,).\nonumber
\ea

The differential cross section is given by
\be
d\sigma = \lim_{t,V\to\infty} \frac{1}{tV}
 \frac{\frac{1}{2}\sum\limits_{\lambda}(T^{\lambda} T^{\lambda
     \dagger})}{v(2E)(2E')(2\nu)(2\omega)}
 \frac{d^3\vec{k}}{(2\pi)^3} \frac{d^3\vec{p}\,'}{(2\pi)^3}
\ee
where $t(time),V(volume)\to\infty$, $v$ is a relative $\gamma A$ velocity. In
the laboratory, that is in the system in which the nucleus is at rest,
$v=1$. The square of the $\delta^4$ functions in $T^{\lambda} T^{\lambda\dagger}$ reads
$
[(2\pi)^4 \delta^4(p-p'+q-k)]^2
= t V (2\pi)^4 \delta^4(p-p'+q-k)
$ where $p=(E,\vec{p}\,)$, $p'=(E',\vec{p}\,')$,
$q=(\nu,\vec{q}\,)$ and $k=(\omega,\vec{k}\,)$.

In the forward kinematics studied here the mass of the nucleus
is considerably larger than the recoiling momentum of
the residual nucleus, {\it i.e.} $E'\to m_A$ and
in the laboratory $\delta(E-E'+\nu-\omega) \to \delta(\nu-\omega)$.
Then the differential cross section  integrated over the three
momentum of the residual nucleus and the energy $\omega$ of the
outgoing meson takes the form
\be
\label{csTOT}
\frac{d\sigma}{d\Omega} = \frac{1}{32\pi^2}\frac{|\vec{k}|}{\nu}
\sum\limits_{\lambda} (M^{\lambda}M^{\lambda\dagger}).
\ee
Here the reduced matrix element $M^{\lambda}$ is given by
\ba
\label{Mreduced}
M^{\lambda} &=& \int \frac{d^4{l}}{(2\pi)^4}
\left[ \sum\limits_{m=\gamma,\omega,\rho}
   d^{(m)}_P({l}\,) \right]  \mathcal{F}_A({l}\,)  \\
&&\times
\varepsilon^{\mu\nu\alpha\beta}q_{\mu}\epsilon_{\nu}^{\lambda}
l_{\alpha}n_{\beta}
\int
d^4{\zeta} e^{-i(l-k+q)\zeta} \chi^{(-)*}_{P}(\vec{\zeta}\,), \nonumber
\ea
where
\be
\label{nf}
n=(1,0,0,0) \simeq  \frac{(p+p')}{\sqrt{2E}\sqrt{2E'}}
\ee
reflects the static approximation for the target nucleus in the laboratory.
It is advantageous to divide Eq.~(\ref{Mreduced}) into two parts
\be
M^{\lambda} = M^{\lambda}_{(1)} + M^{\lambda}_{(2)},
\ee
using
\be
\int
d^4{\zeta} e^{-i(l-k+q)\zeta} \left(1+\left[\chi^{(-)*}_{P}(\vec{\zeta}\,)-1\right]\right).
\ee
The first term here describes a plane wave approximation (PWA). Using
$
\int d^4{\zeta}
e^{-i({l}-{k}+{q}\,){\zeta}} = (2\pi)^4 \delta^4({l}-{k}+{q})
$
the plain wave amplitude takes the from
\be
\label{M1}
M^{\lambda}_{(1)} = \left[ \sum\limits_{m=\gamma,\omega,\rho}
   d^{(m)}_P(\vec{k}-\vec{q}\,) \right]  \mathcal{F}_A(\vec{k}-\vec{q}\,)
\,   \left(\left[\vec{k} \times \vec{q}\,\right] \cdot \vec{\epsilon}^{\lambda}\right).
\ee
If we ignore in Eq.~(\ref{M1}) the $\omega(\rho)$-exchange contributions
then one gets for $\frac{d\sigma}{d\Omega}$ a Primakoff formula, see Eq.~(\ref{PrimStandF}).

The distorted part $M^{\lambda}_{(2)}$ takes into account the FSI of
outgoing mesons. It is given by
\ba
M^{\lambda}_{(2)} &=& \int \frac{d^3\vec{l}}{(2\pi)^3}
\left[ \sum\limits_{m=\gamma,\omega,\rho}
   d^{(m)}_P(\vec{l}\,) \right]  \mathcal{F}_A(\vec{l}\,) \left(
     \left[\vec{l} \times\vec{q}\,\right]\cdot \vec{\epsilon}^{\lambda}\right) \nonumber
 \\
&& \times    \int
d^3\vec{
\zeta}  \, e^{i(\vec{l}-\vec{k}+\vec{q})\vec{\zeta}}\, \left[\chi^{(-)*}_{P}(\vec{\zeta}\,)-1\right].
\ea
For the numerical calculations one can further simplify the distorted amplitude
using the underlying spatial symmetry of the problem. Changing the order of
integrations
\ba
M^{\lambda}_{(2)} &=&  \int
d^3\vec{
\zeta}  \, e^{-i(\vec{k}-\vec{q})\vec{\zeta}} \nonumber \\
&\times& \left(\left[ \frac{1}{i}\vec{\nabla}_{\zeta} V(\vec{\zeta}\,)  \times \vec{q}
\,\right]\cdot \vec{\epsilon}^{\lambda}\right)
\left[\chi^{(-)*}_{P}(\vec{\zeta}\,)-1\right],~~~~
\ea
where
\be
V(\vec{\zeta}\,) = \int \frac{d^3\vec{l}}{(2\pi)^3}
\left[ \sum\limits_{m=\gamma,\omega,\rho}
   d^{(m)}_P(\vec{l}\,) \right]  \mathcal{F}_A(\vec{l}\,) \, e^{i\vec{l}\,\vec{\zeta\,}}.
\ee
Since, $V(\vec{\zeta}\,)=V(|\vec{\zeta}\,|)$ is spherically symmetric its
divergence can be written as
\be
\vec{\nabla}_{\zeta} V(\vec{\zeta}\,)
= \hat{\zeta} \, \frac{\partial
  V(|\vec{\zeta}\,|)}{\partial |\vec{\zeta}\,|},
\ee
where $\hat{\zeta}$ is a unit three vector in the direction of $\vec{\zeta}$.
The resulting amplitude is a spatial integral of the form
\ba
\label{MintSpa}
M^{\lambda}_{(2)} &=&  \int d^3\vec{
\zeta}  \, e^{-i(\vec{k}-\vec{q})\vec{\zeta}} \nonumber \\
&\times& \left(\left[\frac{1}{i}\hat{\zeta} \times \vec{q}  \,\right]\cdot \vec{\epsilon}^{\lambda}\right)
\left[\chi^{(-)*}_{P}(\vec{\zeta}\,)-1\right] \frac{\partial
  V(|\vec{\zeta}\,|)}{\partial |\vec{\zeta}\,|}.~~~~
\ea

At high energies the distortion factor is well described by the eikonal form with
\be
\label{DFmeson}
\chi^{(-)*}_{P}(\vec{\zeta}\,) = e^{
- \frac{1}{2}  \sigma_{P}
\int\limits_{\zeta_z}^{\infty} \rho_A(\vec{\zeta}_{\perp},\hat{k} z') dz'
}
\ee
where $\hat{k}$ is a unit three vector in the direction of outgoing meson,
$\sigma_{P}$ stands for the $PN$ total cross section and
$\rho_A$ denotes the nuclear density, $\int d^3\vec{\zeta}\rho_A(\vec{\zeta}\,)=A$.

At extreme forward angles studied here the dependence on $\hat{k}$ in
Eq.~(\ref{DFmeson}) can be readily neglected and the $z'$ integration runs along the
beam axis. In this case the azimuthal angle in Eq.~(\ref{MintSpa}) can be further integrated out analytically. The final result is given
by the following expression ($\delta=|\vec{l}\,|$, $b=|\vec{\zeta}_{\perp}|$ and $z=\zeta_z$)
\ba
\label{M2FSIfinal}
&& M^{\lambda}_{(2)} =
\left(\left[\vec{k}
  \times\vec{q}\,\right]\cdot \vec{\epsilon}^{\lambda}\right)
\int\limits_{-\infty}^{\infty} d z
\int\limits_{0}^{\infty} d b \frac{2\pi b^2}{\sqrt{b^2+z^2} }  \,
 \nonumber \\
&& \times
e^{-i(|\vec{k}\,|\cos\theta-\nu)z}  \frac{J_1\left(|\vec{k}\,|b \sin\theta\right)}{|\vec{k}\,| \sin\theta}\left[\chi^{(-)*}_{P}(b,z)-1\right]   \\
&& \times
\int\limits_{0}^{\infty} \frac{d\delta \delta^3}{2\pi^2}
\left[ \sum\limits_{m=\gamma,\omega,\rho}
   d^{(m)}_P(\delta) \right]  \mathcal{F}_A(\delta) \,
j_1\left(\delta \,\sqrt{b^2+z^2}\right). \nonumber
\ea
Here $J_1$ denotes the (cylindrical) Bessel function of the first kind, $j_1$
is the spherical Bessel function,
$|\vec{k}\,|\cos\theta-\nu$ and $|\vec{k}\,|\sin\theta$ describe, respectively, the
longitudinal and transverse momentum transfers to the nucleus.
We note that Eqs.~(\ref{M1}) and~(\ref{M2FSIfinal}) contain both the Primakoff amplitude and the nuclear coherent amplitude as special cases.

\subsection{Initial state interactions}
In a VDM followed here the $\gamma$ converts into the virtual vector mesons
$V=\rho^0,\omega$ and $\phi$. According to the uncertainty principle, these can
further fluctuate into real state
which then  may rescatter coherently inside the nucleus before the
production point. In this case the eikonal wave function of a photon
is given by~\cite{Falter:2002vr,Bauer:1977iq}
\be
\label{Phwf}
\Phi^{(+)}_{\gamma}(\zeta)= e^{-i\nu t}
\varphi^{(+)}_{\gamma}(\vec{\zeta}\,) = e^{-iq\zeta}
 \Big(1 -  S_V(\vec{\zeta}\,)\Big),
\ee
where $S_V(\vec{\zeta\,})$ denotes the shadowing term ($\vec{\zeta\,}=(\vec{\zeta}_{\perp},z)$)
\ba
\label{S_V}
S_V(\vec{\zeta}\,) =
\int\limits_{-\infty}^{z} dz_i \rho_A(\vec{\zeta}_{\perp},z_i)
  \frac{\sigma_{V}}{2}(1-i\beta_{V})
   e^{iq_{V}(z_i-z)} \nonumber
     \\  \times
e^{-\frac{\sigma_{V
      }}{2} (1-i\beta_{V})
       \int\limits_{z_i}^{z} dz_{j}
    \rho_A(\vec{\zeta}_{\perp},z_j)}.~~~
\ea
Here the distance $l_{V}$ that the photon travels as a hadron is defined as
$l_{V} =|q_{V}^{-1}|$ where
$q_{V}= \nu -\sqrt{\nu^2-m_{V}^2}$.
For the average photon energy $\nu\simeq 5.2$~GeV in the PRIMEX experiment~\cite{PRIMEX}
the coherence length, {\it e.g.} $V=\rho^0(770)$, is $l_{\rho} \simeq
3.5~fm$. This value is well within
the dimension of a nucleus and is just the in-medium mean free path of
$\rho$. Therefore, we expect sizable shadowing corrections at JLAB.
In Eq.~(\ref{S_V}) different VDM components shadow according to the corresponding $V$-nucleon cross
sections $\sigma_{V}$ and $\beta_V=\mathfrak{R}f_V(0)/\mathfrak{I}f_V(0)$
where $\mathfrak{R}f_V(0)(\mathfrak{I}f_V(0))$ describes the real (imaginary) part of
the forward $VN\to VN$ scattering amplitude $f_V(0)$.

The anomalous VDM interactions involve the derivatives of vector
fields, see Eq.~(\ref{L_VVM_f}). Therefore, in the shadowed amplitude one has to keep the
derivative of the incoming distorted wave
\be
\label{Mph1}
M^{\lambda} \propto   [\,\vec{l}   \times \vec{\nabla}\varphi^{(+)}_{\gamma}  \,
  ] \cdot \vec{\epsilon}^{\lambda} = \varepsilon_{ijk}
\epsilon^{\lambda}_i l_j  \partial_k \varphi^{(+)}_{\gamma},
\ee
where
$\vec{\nabla}\varphi^{(+)}_{\gamma}$ picks up the components
($\varepsilon_{ijk}$ is antisymmetric) which do not lie along the vector
$\vec{l}$, {\it i.e.}, $k \neq j$.  The
electromagnetic current conservation imposes further constraints on the
derivative in (\ref{Mph1}). The amplitude, Eq.~(\ref{Mreduced}), can be written as
$M^{\lambda} = \mathcal{J}^{\nu} \epsilon_{\nu}^{\lambda}$ where by gauge invariance the current
$\mathcal{J}^{\nu}$ is conserved $\mathcal{J}^{\nu}q_{\nu}=0$. In the static
limit, see Eq.~(\ref{nf}), the time component $\mathcal{J}^0=0$. Therefore, the product of
the vector part of the current and $\vec{q}$ should give
\be
\label{Mph2}
[\, \vec{l}  \times   \vec{\nabla}\varphi^{(+)}_{\gamma} \,
  ] \cdot \vec{q} = \varepsilon_{ijk}
q_{i} l_j  \partial_{k} \varphi^{(+)}_{\gamma}  = 0.
\ee
Because of Eq.~(\ref{Mph1}),  this is
satisfied when $\vec{\nabla}\varphi^{(+)}_{\gamma}$ picks up a term which is
parallel to $\vec{q}=(0,0,\nu)$.
Thus, in the derivative we need the longitudinal part
$
\partial_z \varphi^{(+)}_{\gamma}
$ only.

The distortion factor of a photon $\chi^{(+)}_{V}$ takes the form
\ba
\label{DFphoton}
\chi^{(+)}_{V}(\vec{\zeta}\,) =  \hspace{6.7cm} \\ \left[1+i
\frac{\sigma_{V}}{2\nu}(1-i\beta_{V})
\rho_A(\vec{\zeta}\,)\right]
\Big(1 - S_V(\vec{\zeta}\,)\Big)
+ \frac{q_V}{\nu}
S_V(\vec{\zeta}\,). \nonumber
\ea
In the strong production of $\pi^0,\eta$ and $\eta'$ only $V=\rho^0$ and $\omega$
matter. Recall that by strict OZI rule we neglected the exchange of $\phi$. From the fit to data $\sigma_{\rho}\simeq
\sigma_{\omega}$~\cite{Bauer:1977iq} and the common shadowing factor factorizes in
the production point. In the Primakoff production of $\eta$ and $\eta'$  we have an
additional $\phi$ component. However, the strength of the $\gamma\gamma^*\to\phi\gamma^*\to\eta(\eta')$ transition is much
smaller than the conversions induced by the $\rho^0$ intermediate state.
Although, $\sigma_{\phi}<\sigma_{\rho}(\sigma_{\omega})$, without much impact
  on the results we may assume $\sigma_{\rho}\simeq\sigma_{\omega} \simeq \sigma_{\phi}$
and further use the same
shadowing corrections for all $\gamma-V$ transitions.

Then the distorted part of the amplitude which takes into
account both the shadowing and meson FSI reads
\ba
\label{M2shad}
M^{\lambda}_{(2)} &=&  \int \frac{d^3\vec{l}}{(2\pi)^3}
\left[ \sum\limits_{m=\gamma,\omega,\rho}
   d^{(m)}_P(\vec{l}\,) \right]  \mathcal{F}_A(\vec{l}\,)
\left(
     \left[\vec{l} \times\vec{q}\,\right]\cdot \vec{\epsilon}^{\lambda}\right)
 \nonumber
 \\
&& \times    \int
d^3\vec{\zeta}  \, e^{i(\vec{l}-\vec{k}+\vec{q})\vec{\zeta}}\,
[\chi^{(-)*}_{P}(\vec{\zeta}\,)\chi^{(+)}_{V}(\vec{\zeta}\,)-1].
\ea
Using the same steps as before one arrives at an expression similar to
Eq.~(\ref{M2FSIfinal}) with the replacement
\be
[\chi^{(-)*}_{P}(b,z)-1] \to [\chi^{(-)*}_{P}(b,z)\chi^{(+)}_{V}(b,z)-1].
\ee

\section{Results}
In this section we compare the model results with $\pi^0$ photoproduction
data measured at JLAB~\cite{PRIMEX}. Then we extend the results to the
$\eta$ and $\eta'$ channels. At first, we specify the input for the nuclear
form factors and meson-nucleon cross sections in FSI and ISI.

The nuclear form factors entering the amplitudes are defined as follows
\ba
\mathcal{F}_A(|\vec{l}\,|) &=&\frac{1}{Z} \int d^4 x e^{ilx} \delta(x_0)
\rho_Z(\vec{x}) \\
&& = \frac{4\pi}{Z} \int d|\vec{x}\,|\,|\vec{x}\,|^2
j_0(|\vec{l}\,||\vec{x}\,|) \rho_Z(|\vec{x}|), \nonumber
\ea
where the charge  density distributions $\rho_Z(\vec{x})=\rho_Z(|\vec{x}|)$ of nuclei are parameterized according to
Ref.~\cite{ADNDT}. For the matter
density distributions we assume $\rho_A(\vec{x})=\frac{A}{Z}\rho_Z(\vec{x})$.

\begin{figure}[b]
\begin{center}
\includegraphics[clip=true,width=1\columnwidth,angle=0.]
{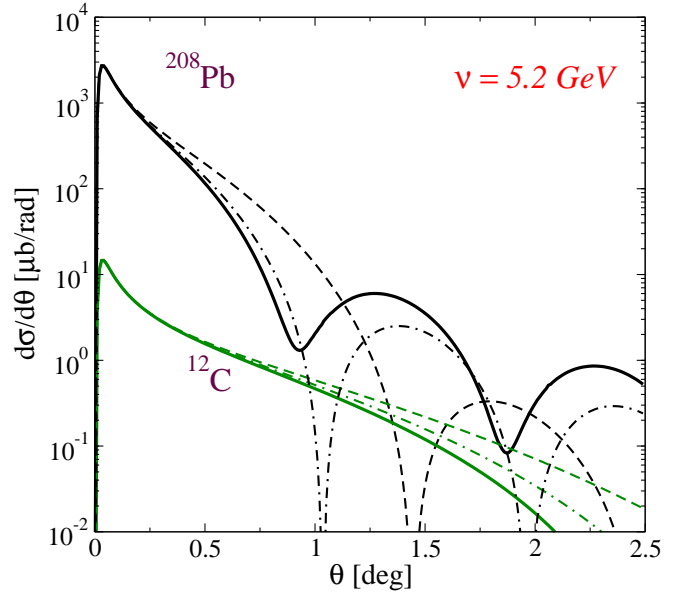}
\caption{\label{Coulomb.Pb.C}
Primakoff effect in coherent photoproduction of $\pi^0$-meson off $^{12}$C (bottom curves) and $^{208}$Pb
(top curves) nuclei in the kinematics of the PRIMEX experiment~\cite{PRIMEX}. The
incoming photon energy
$\nu=5.2$~GeV in the laboratory. The dashed curves describe the calculations
without any in-medium interactions of incoming ($\gamma$) and outgoing $(\pi^0)$ particles. The dash-dotted
curves describe the effect of FSI. The solid curves include FSI
and $\gamma$-shadowing in ISI.
\vspace{-0.3cm}
}
\end{center}
\end{figure}

The $\rho^0$-nucleon cross section $\sigma_V=\sigma_{\rho}$ and
$\beta_{V}=\beta_{\rho}$  in Eqs.~(\ref{S_V}) and (\ref{DFphoton})
are taken from~\cite{Bauer:1977iq}.
By isospin the total $\pi^0 N$ cross section in Eq.~(\ref{DFmeson}) would be
$\sigma_{\pi^0}=\frac{1}{2}(\sigma_{\pi^+}+\sigma_{\pi^-})$. This value is in
agreement with the quark model estimate
$\sigma_{\pi^0}\simeq\sigma_{\rho}$. For
the $\eta N$ and $\eta' N$ cross sections $\sigma_{\eta}$ and $\sigma_{\eta'}$,
respectively, we assume the same value as $\sigma_{\pi^0}(\sigma_{\rho})$. Then the only model input is
$\sigma_V=\sigma_{\rho}$ from the fit to data by~\cite{Bauer:1977iq} where
$\sigma_{V}=20.8\left(1+\frac{0.766}{\sqrt{\nu^2-m_{\rho}^2}}\right)$~mb and
$\beta_{V}=\frac{0.766}{\sqrt{\nu^2-m_{\rho}^2}+0.766}$.

\begin{figure*}[ht]
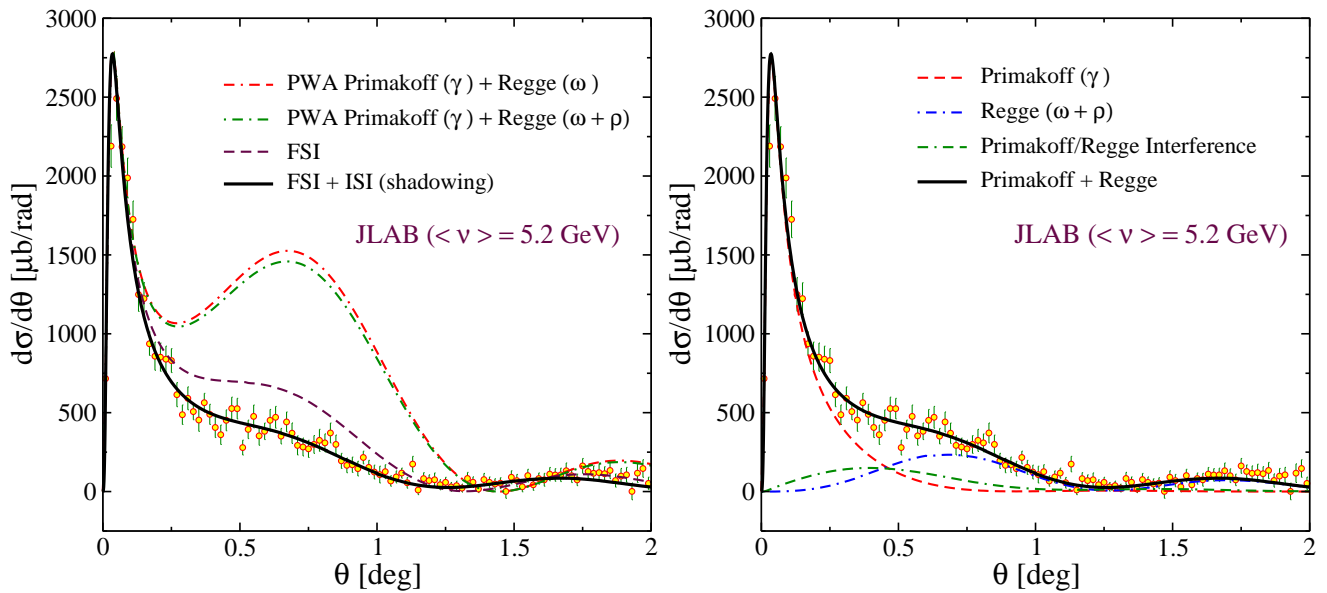

\begin{center}
\includegraphics[clip=true,width=1\columnwidth,angle=0.]
{JLAB.Pb.V2.eps}
\includegraphics[clip=true,width=1\columnwidth,angle=0.]
{JLAB.Pb.eps}
\caption{\label{Prim.JLAB.Pb}
Differential cross section ${d\sigma}/{d\theta}$ in the reaction
$(\gamma,\pi^0)$ off $^{208}$Pb target as a function of $\pi^0$
production angle $\theta$ in the laboratory. The experimental data
are from Ref.~\cite{PRIMEX}.
The solid curves in the left and right panels describe the model coherent cross section
with meson FSI and photon shadowing in ISI with the average value of $\nu=5.2$~GeV. The
Primakoff $(\gamma)$ and nuclear coherent (Regge $(\omega+\rho)$ exchange)  contributions are
taken into account.
Left panel: The dash-dotted curve describe the model
results in a plain wave approximation. The dash-dash-dotted curve
corresponds to the plain wave approximation (PWA) without the exchange of $\rho$-Regge
trajectory. The dashed curve accounts for the effect of FSI only.
Right panel:
The dashed and dash-dotted curves describe the nuclear coherent Primakoff and Regge ($\omega+\rho$)
exchange contributions, respectively. The dash-dash-dotted curve describes the
interference between the Primakoff and Regge exchange contributions.}
\vspace{-0.3cm}
\end{center}
\end{figure*}

\subsection{$\pi^0$ photoproduction at JLAB}
High precision measurements of the differential cross sections
$d\sigma/d\theta$ in $\pi^0$ photoproduction off nuclei at forward angles have been
performed in the PRIMEX experiment at JLAB~\cite{PRIMEX} with $^{12}$C and
$^{208}$Pb as the targets. The $\pi^0\to\gamma\gamma$ decay width has been
extracted  with the
magnitudes of nuclear coherent, nuclear incoherent and the phase angle between
the Primakoff and the nuclear coherent amplitudes being parameters in the extraction.

At first, we consider the effects of FSI and shadowing on the Primakoff
effect. In Fig.~\ref{Coulomb.Pb.C} we show the in-medium distortion of
the Primakoff signal in the coherent production of $\pi^0$ off $^{12}$C
(bottom curves) and $^{208}$Pb (top curves) nuclei in the kinematics of
the PRIMEX experiment. The incoming photon energy in the laboratory is
$\nu=5.2$~GeV; this is an average value used at JLAB. The dashed curves
describe the calculations without any in-medium interactions of incoming
($\gamma$) and outgoing $(\pi^0)$ particles. The dash-dotted curves include
the effect of FSI only. The solid curves include FSI of pions and photon
shadowing in ISI.  As one can see, the latter have practically no influence
on the Coulomb peak at forward angles for both nuclei where the width
$\Gamma_{\pi^0\to\gamma\gamma}$ is extracted. For $^{208}$Pb the tail of the
Primakoff signal at $\theta \sim 0.5^{\circ}-0.75^{\circ}$ gets largely distorted by FSI,
but the magnitude of the signal is already very small there.

\begin{figure*}[t]
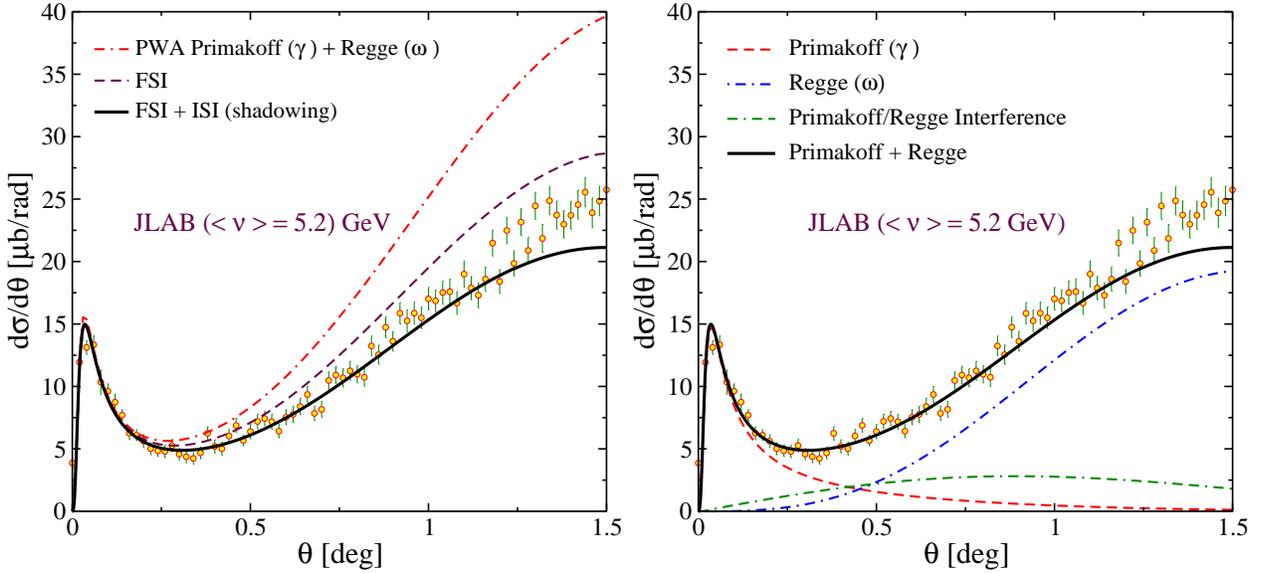

\begin{center}
\includegraphics[clip=true,width=0.95\columnwidth,angle=0.]
{JLAB.Carbon.V2.eps}
\includegraphics[clip=true,width=0.95\columnwidth,angle=0.]
{JLAB.Carbon.eps}
\caption{\label{Prim12CJLAB}
Differential cross section ${d\sigma}/{d\theta}$ in the reaction
$(\gamma,\pi^0)$ off $^{12}$C target as a function of $\pi^0$
production angle $\theta$ in the laboratory. The experimental data
are from Ref.~\cite{PRIMEX}. The notations for the curves in both
panels are the same as in Fig.~\ref{Prim.JLAB.Pb}.
\vspace{-0.3cm}
}
\end{center}
\end{figure*}

\begin{figure}[b]
\begin{center}
\includegraphics[clip=true,width=1.\columnwidth,angle=0.]
{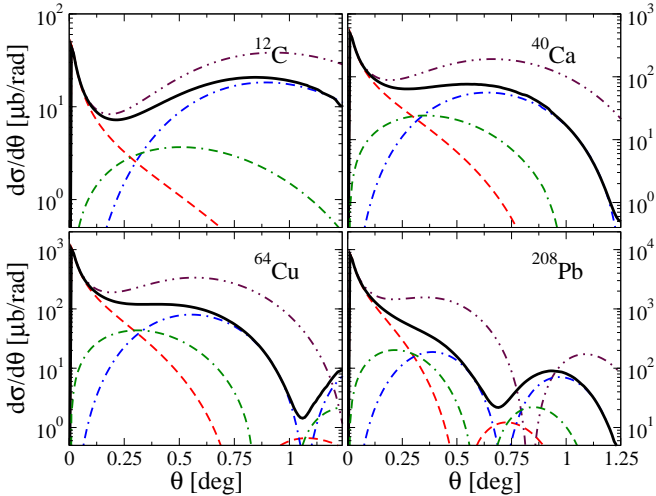}
\caption{\label{Pi.A.9GeV}
Differential cross section $d\sigma/d\theta$ in the reaction $A(\gamma,\pi^0)A$
off $^{12}$C, $^{40}$Ca, $^{64}$Cu and $^{208}$Pb nuclei. The incoming photon energy in
the laboratory is $\nu=9$~GeV. The solid curves describe the model results.
The dashed, dash-dotted and dash-dash-dotted curves describe the Primakoff,
Regge and interference Primakoff/Regge cross sections, respectively. The
dot-dot-dashed curves correspond to the PWA.
\vspace{-0.0cm}
}
\end{center}
\end{figure}

\begin{figure*}[htb]
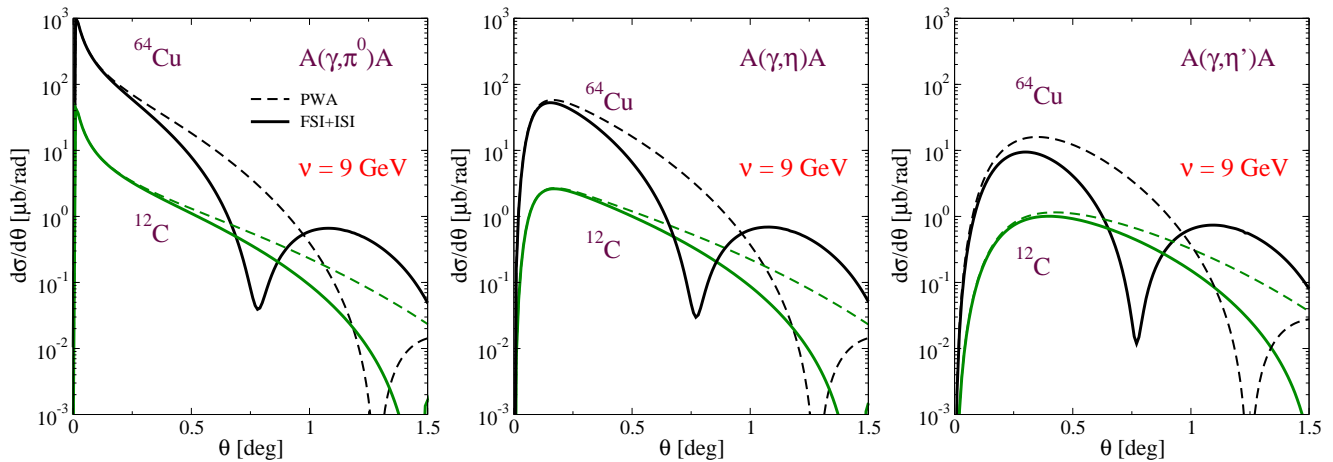

\begin{center}
\includegraphics[clip=true,width=0.665\columnwidth,angle=0.]
{Coulomb.Pi.eps}
\includegraphics[clip=true,width=0.665\columnwidth,angle=0.]
{Coulomb.Eta.eps}
\includegraphics[clip=true,width=0.665\columnwidth,angle=0.]
{Coulomb.EtaPr.eps}
\caption{\label{Coulomb.9GeV}
Primakoff effect in the coherent production of $\pi^0$ (left panel), $\eta$
(middle panel) and $\eta'$ (right panel) off $^{12}$C (bottom curves) and $^{64}$Cu
nuclei. The calculations are performed for the photon energy $\nu=9$~GeV in the laboratory.
The solid (dashed)  curves describe the calculations with (without) final and
initial state interactions.
\vspace{-0.5cm}
}
\end{center}
\end{figure*}

To compare the model results with experimental data the theoretical cross sections have
to be folded with the photon energy spectrum and the angular resolution.
The experimental spectrum of the incident photons is essentially an uniform
distribution over the tagged range of $\nu =
4.9-5.5$~GeV~\cite{PhDThesisPrim}. The angular distribution to be compared with data is
\be
\frac{1}{\Delta \nu}
\int\limits_{\Delta\nu} d\nu
\int d{\theta_i}  \,
  \frac{d\sigma(\nu,\theta_i)}{d\theta_i}  P(\theta,\theta_i)
\ee
where $P(\theta,\theta_i)=\frac{1}{b\sqrt{\pi}}e^{-(\theta-\theta_i)^2/b^2}$
with $b=\frac{\Delta\theta}{2\sqrt{\ln 2}}$ describes the Gaussian angular resolution with the width of
$\Delta \theta = 0.4$~mrad~\cite{AG}.

In Fig.~\ref{Prim.JLAB.Pb} the experimental differential cross
section~\cite{PRIMEX} $d\sigma/d\theta$ in the reaction
$(\gamma,\pi^0)$ off $^{208}$Pb is compared with our model results.
The data exhibit the sharp forward peak characteristic of Primakoff
production. The weaker dependence at larger $\theta$ is due to
strong coherent $\pi^0$ production
\footnote{Note that the experimental spectrum contains both the
coherent and incoherent cross sections.}.
The incoherent cross section for a $^{208}$Pb target (not shown here) is tiny at these small
angles and thus makes
no  significant contribution in the region of the Primakoff
signal~\cite{PRIMEX}. The solid curves in the left and right panels
of Fig.~\ref{Prim.JLAB.Pb} are our model results for the total
coherent cross section. The Primakoff ($\gamma$) and the nuclear coherent 
Regge $\omega$ and $\rho$ exchange contributions together with FSI and ISI are taken into account. In the left panel the dash-dotted curve describes the model results in a PWA.
Also the dash-dash-dotted curve describes the plain wave results without
the exchange of the $\rho$-meson Regge trajectory. Because of the large
excess of neutrons over the number of protons in $^{208}$Pb the
$\rho$-exchange contributes destructively in this case.
The bump between $\theta \sim 0.5^{\circ}-1^{\circ}$ is due to nuclear coherent
production. It gets strongly reduced by FSI of the
produced pions. In the left panel the dashed curve accounts for FSI of pions
without shadowing in ISI.

In the right panel of Fig.~\ref{Prim.JLAB.Pb} we show different reaction
mechanisms which contribute to the cross section. The dashed curve is the
Primakoff effect. The dash-dotted curve describe the Regge exchange contributions, labeled
as nuclear coherent in Ref.\ \cite{PRIMEX}.
The dash-dash-dotted curve describes the interference between the
Primakoff and Regge exchange amplitudes. 

In Fig.~\ref{Prim12CJLAB} we compare the model results with the measured
differential cross sections ${d\sigma}/{d\theta}$ in the $\pi^0$
phototoproduction off $^{12}$C nucleus. The notations for the curves in
both panels are the same as in Fig.~\ref{Prim.JLAB.Pb}. In the production
of $^{12}$C the isospin filtering is realized exactly. Therefore, there
is no $\rho$-exchange contribution in this case. The nuclear coherent
background (dashed curve in the right panel) is dominated by
the exchange of the $\omega$-Regge trajectory. The model results
(solid curves) are in agreement with data below $\theta\simeq 1^{\circ}$.
For $^{12}$C this is a region where the coherent Primakoff and Regge
exchange contributions dominate the cross section. The incoherent cross
section is largely suppressed in this region. However, it gives a sizable
contribution beyond that region where the difference between the
model results and data is due to the incoherent
$\pi^0$ production~\cite{PRIMEX}.

Note that the calculation of the incoherent cross section is conceptually
different. Since the nucleus breaks up (or gets excited) in this case the coherence
is lost and the nuclear cross section is given by the sum of individual
nucleon cross sections corrected for the effects of Fermi motion, Pauli
blocking, shadowing and FSI. The FSI contain coupled channel effects, such a
charge exchange, so that the final $\pi^0$ could not be the one initially
produced. In line with our transport theoretical
model~\cite{Kaskulov:2008ej} these calculations will be presented in
forthcoming publications.

Summarizing these comparisons, the overall agreement of the full
calculation with present data is remarkable and there is no need to introduce any
further fit parameters to describe the data at forward angles. Note that
contrary to the treatment in~\cite{Gevorkyan:2009ge,PRIMEX} the relative weights of
the various contributions are all fixed from the elementary process so that no
further adjustment to the nuclear data can be performed. We recall that in \cite{PRIMEX} the relative strengths and phases of the various amplitudes were fitted to experiment and turned out to depend even on mass number.

This agreement encourages us now to make predictions for higher energies and other mesons. The model predictions for higher energies ($\nu=9$~GeV) are shown in
Fig.~\ref{Pi.A.9GeV}. The solid curves describe the model results for
the coherent cross section. The dashed, dash-dotted and dash-dash-dotted
curves correspond to the Primakoff, Regge and interference Primakoff/Regge
cross sections, respectively. The dot-dot-dashed curves correspond to the plane
wave approximation. As one can see, in all cases the Primakoff peak
dominates over the Regge behavior of the nuclear background. Furthermore, for heavy nuclei
the Primakoff effect dominates the overall reaction mechanism.

The in-medium distortion of the Primakoff signal for pions at $\nu=9$~GeV off $^{12}$C (bottom curves) and $^{64}$Cu nuclei is shown in the left panel of Fig.~\ref{Coulomb.9GeV}.
In the middle and right panel we also present the results for the Primakoff effect in the coherent production of $\eta$  and $\eta'$ mesons, respectively. In Fig.~\ref{Coulomb.9GeV} the dashed curves describe the bare Coulomb component in a plane wave approximation. The solid curves include FSI and shadowing corrections. As one can see, in the $\pi^0$ production (left panel) at the maximum of the Primakoff effect the in-medium distortion of incoming and outgoing waves does not affect the $\gamma\gamma^*\to\pi^0$ signal. In the off-forward region, however, the distortion is large and changes the shape of the signal. These changes are much more pronounced for heavy systems.

In contrast, the peak positions for the $\eta$ and $\eta'$ signals are shifted to larger
angles because at fixed photon energy the position of the Primakoff signal
depends on the mass of the produced meson $\propto m_P^2/\nu$.  Since the photon excitation to the heavy $\eta$ and $\eta'$ states requires somewhat bigger momentum transfers to the nucleus than for the pions the range of the corresponding interactions is much smaller. These conversions are then not peripheral anymore as in the $\pi^0$ case and may proceed deep inside the nucleus. As a consequence, Primakoff and nuclear coherent excitations become entangled.

\subsection{$\eta(547)$ photoproduction off nuclei}

In Fig.~\ref{PrimEtaAdep5} we present our results for the $\eta$
production in the kinematics accessible at present JLAB energies.
The calculations correspond to the incoming photon energy
$\nu=5.2$~GeV in the laboratory. Recall that in the elementary production of
nucleons the Primakoff signal is barely visible on top of the strong
background. On the contrary, the filtering of the $\rho$-exchange
makes the nuclear Primakoff effect pronounced
already at these photon energies. For instance, on heavy targets the Coulomb peak
dominates over the Regge exchange contributions. This makes further
experimental studies of the Primakoff effect at present JLAB energies promising.
The situation would be much better at higher energies.

\begin{figure}[t]
\begin{center}
\includegraphics[clip=true,width=1\columnwidth,angle=0.]
{Eta.A.dep.5GeV.eps}
\caption{\label{PrimEtaAdep5}
Differential cross section ${d\sigma}/{d\theta}$ in the
 reaction $A(\gamma,\eta)A$ at $\nu=5.2$~GeV
in the laboratory. The notations for the curves are the same as in
Fig.~\ref{Pi.A.9GeV}.
} \vspace{-0.4cm}
\end{center}
\end{figure}

\begin{figure}[b]
\begin{center}
\includegraphics[clip=true,width=0.92\columnwidth,angle=0.]
{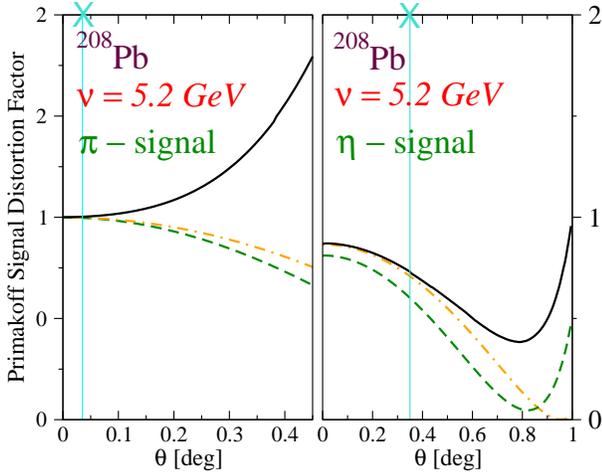}
\caption{\label{SignalDist}
Distortion of the bare Primakoff signal at $\nu=5.2$~GeV in the
 coherent reactions $^{208}$Pb$(\gamma,\pi^0)^{208}$Pb (left panel),
$^{208}$Pb$(\gamma,\eta)^{208}$Pb (right panel) due to FSI (dash-dotted), FSI+ISI
 (dashed) and strong production (solid). The crossed symbol shows the position
 of the maximum of the Primakoff signal where the decay width is extracted.
\vspace{-0.3cm}
}
\end{center}
\end{figure}

\begin{figure}[t]
\begin{center}
\includegraphics[clip=true,width=1\columnwidth,angle=0.]
{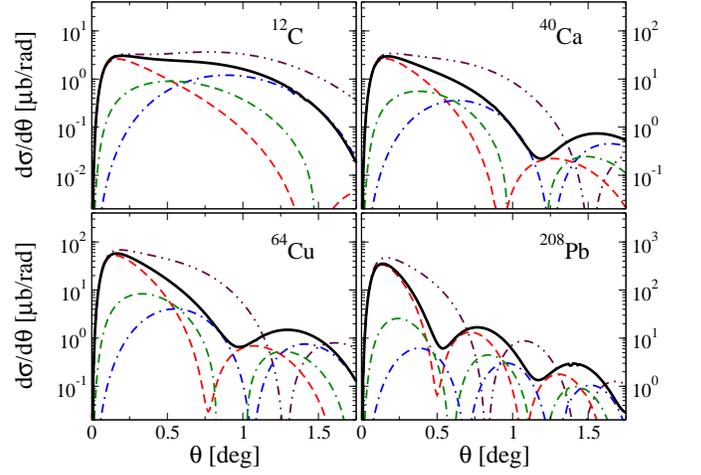}
\caption{\label{PrimEtaAdep9}
Differential cross section ${d\sigma}/{d\theta}$ in the
 reaction $A(\gamma,\eta)A$ at $\nu=9$~GeV
in the laboratory. The notations for the curves are the same as in Fig.~\ref{Pi.A.9GeV}.}
\vspace{-0.4cm}
\end{center}
\end{figure}

For the measurements of the $\pi^0\to\gamma\gamma$ and $\eta \to \gamma\gamma$
decay widths in the Primakoff effect at present JLAB energies,
Fig.~\ref{SignalDist} (left panel)
demonstrates the distortion of the Primakoff signal in the $\pi^0$
photoproduction off $^{208}$Pb target. The distortion factor
in Fig.~\ref{SignalDist} is defined as the ratio of the model
and bare (without FSI and ISI) Primakoff cross sections. As one can see the
distortion of the signal due
to the FSI (dash-dotted), FSI+ISI (dashed) and strong production (solid) is
less than $<1\%$ around the maximum of the Primakoff signal (crossed symbol). On the contrary,
for the $\eta$ (right panel in Fig.~\ref{SignalDist}) the distortion is already significantly larger.

The results for the Primakoff effect in the photoproduction of $\eta$ mesons
at $\nu=9$~GeV are
shown in the middle panel of Fig.~\ref{Coulomb.9GeV}. This energy region corresponds to
an approved experimental proposal for the $\eta$ production via the Primakoff
effect in Hall D at JLAB~\cite{etaGasp}.
As one can see, the magnitude of the Coulomb signal is barely affected by FSI
and ISI at
forward angles in both
light and heavy systems. In Fig.~\ref{PrimEtaAdep9}
we show the model predictions at $\nu=9$~GeV which include both the Primakoff and
Regge exchange contributions.

\subsection{$\eta'(958)$ photoproduction off nuclei}
In Fig.~\ref{Coulomb.9GeV} (right panel) the results for the Primakoff production of
$\eta'$ meson at $\nu=9$~GeV are shown.
This result is instructive since it suggest
a use of light systems to minimize the in-medium effects in the region of the
Coulomb peak. For heavy systems the signal is strongly affected by the in-medium
interactions. Therefore, the use of light nuclear targets preferable. The price at these energies is smaller cross sections than in
the production off heavy targets.

\begin{figure}[t]
\begin{center}
\includegraphics[clip=true,width=1\columnwidth,angle=0.]
{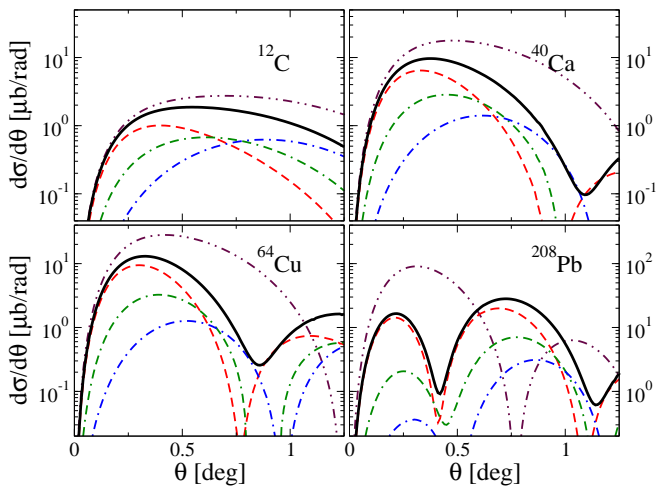}
\caption{\label{PrimEtaPrAdep}
Differential cross section ${d\sigma}/{d\theta}$ in the
 reaction $A(\gamma,\eta')A$ at $\nu=9$~GeV
in the laboratory. The notations for the curves are the same as in Fig.~\ref{Pi.A.9GeV}.
\vspace{-0.2cm}
}
\end{center}
\end{figure}

The coherent production of $\eta'$ for
$\nu=9$~GeV are shown in Fig.~\ref{PrimEtaPrAdep}. The
notations for the curves are the same as in Fig.~\ref{Pi.A.9GeV}.
Because of the suppression of the $\rho$-Regge exchange the Primakoff signal
is large (dashed curves) and dominates over the nuclear coherent background
(dash-dotted curves). The relative contribution of the nuclear coherent
cross section is getting small with increasing mass number of a nucleus.
However, the interference cross section (dash-dash-dotted curves) gets large.
This feature can be seen in the $\eta$ production also. Interestingly what
contaminates the Primakoff signal is a large interference between the
Coulomb and strong amplitudes.

\section{Summary}
In summary, we have considered the high energy coherent photoproduction of pseudoscalar
mesons $\pi^0,\eta$ and $\eta'$ off nuclei. All the calculations presented
here have been performed at extreme forward angles where the Primakoff effect is
expected to dominate the nuclear photoproduction mechanism.
At high energies the Primakoff signal is expected to be well separated from
the nuclear background contributions. If true this allows to measure the
radiative $\gamma\gamma$ decay widths of pseudoscalar mesons. 

In reality the
interference of the Primakoff amplitude with the nuclear coherent amplitude
contaminates the signal. Furthermore, the conversion of photons to mesons
deep inside the nucleus distorts the signal by FSI and photon shadowing in ISI.
The data analysis and width extraction thus have to be based on a model
which is able to describe the different production mechanisms reliably.


In the present work we have proposed such a model for the coherent
production of mesons which treats the Coulomb and strong nuclear
components in the same footing. Starting from the VDM description of the
anomalous sector we have established the transition amplitudes which have
been used to describe the photoproduction process off nucleons.
A use of Reggeon exchanges allows to start from the description of the
elementary cross sections off nucleons and take into account the complex
phase between the Primakoff and Regge exchange amplitudes. In VDM both
the radiative $\gamma\gamma\to P$ and $\gamma V \to P$ transitions are
correlated. This allows to calculate consistently the in-medium
interactions of mesons and photons in the final and initial states,
respectively. Contrary to previously existing approaches our model describes simultaneously the Primakoff  $(\gamma)$, Regge $(\omega,\rho)$ and the interference Primakoff/Regge exchange
contributions with the same input from the reactions off \emph{nucleons}.

Furthermore, we have explicitly demonstrated the importance of FSI and shadowing
corrections in ISI in the coherent $\pi^0$ photoproduction reactions
of \emph{nuclei}. The model describes the forward region around the Primakoff
signal measured in the $\pi^0$ photoproduction at JLAB very well.
There was no need to introduce any additional 
parameters to match the model cross sections with the experimental data.

The extremely small distortion of the $\pi^0$ Primakoff signal by other
processes explains why the various nuclear fit parameters in the
analysis of~\cite{Gevorkyan:2009ge} do not affect the actual signal. This is
very different, however, for the $\eta$ and $\eta'$ Primakoff experiments
where the strong nuclear distortions are large in the peak region and
have to be well under control. 
Therefore, we have also calculated cross sections for the photoproduction of
$\eta$ and $\eta'$ mesons. We find that because of the isospin filtering of the
$\rho$-exchange the Primakoff signal in the $\eta$
production off nuclei is still large compared with the nuclear coherent background already at present JLAB energies. However, at these energies there are already
significant distortions due to FSI, ISI and the strong amplitude, see Fig.~\ref{SignalDist} (right panel).
In the case of $\eta$ and $\eta'$  high energy data from future JLAB
experiments may make the study of the Primakoff effect and an
extraction of the $\eta(\eta')\to\gamma\gamma$ decay width possible.  In both $\eta$
and $\eta'$ cases the coherent production
provide at the same time a filter and an amplifying device which allows to
isolate the Primakoff $\gamma$-exchange production mechanisms which otherwise
remain hidden in the background. The observed sensitivity of the
cross sections to the in-medium $\eta$ and $\eta'$ interactions may also provide a
complementary way to measure the unknown $\eta(\eta')$-nucleon cross sections.

\begin{acknowledgments}
We are grateful to A.~Bernstein and A.~Gasparian
for helpful communications and reading the manuscript.

This work was supported by DFG through TR16 and by BMBF.
\end{acknowledgments}

\begin{appendix}
\section{Notations for $P$ and $V_{\mu}$ matrices}
The mixing pattern of the SU(3) singlet $V_0^{\mu}$ and the isospin $I=0$
octet of vector mesons
$V_8^{\mu}$ is supposed to be ideal, that is
$
\omega(782) = \sqrt{\frac{2}{3}} V_0 + \sqrt{\frac{1}{3}} V_8^{I=0}$ and
$\phi(1020)  = \sqrt{\frac{1}{3}} V_0 - \sqrt{\frac{2}{3}} V_8^{I=0}$.
Thus, the vector mesons are described
by the matrix $V_{\mu}$
\be
\left(
\begin{array}{ccc}
\rho^0+\omega     &    0   & 0 \\
   0  & - \rho^0 + \omega & 0 \\
0 & 0 & \sqrt{2}\phi
\end{array}
\right)_{\mu}.
\ee
The diagonal components of the nonet of
pseudoscalar mesons $P=P_8+P_0$ are incorporated into the matrix
\be
\left(
\begin{array}{ccc}
\pi^0+\frac{\eta_8}{\sqrt{3}}+\sqrt{\frac{2}{3}} \eta_0     &   0  &
0 \\
  0   & - \pi^0 +\frac{\eta_8}{\sqrt{3}}+\sqrt{\frac{2}{3}}\eta_0 & 0\\
0 & 0 & -\frac{2\eta_8}{\sqrt{3}} +\sqrt{\frac{2}{3}}\eta_0
\end{array}
\right)
\ee
where $\eta_0-\eta_8$ mixing is described in Sec.~II.
\end{appendix}

\end{document}